\documentclass[nofootinbib,aps,11pt,prd]{revtex4-1}

\usepackage{graphicx,color,verbatim,acronym,epsfig,subfigure,slashed,skak,geometry}
\usepackage{latexsym,amsfonts,amssymb,amsmath,makeidx,mathrsfs,multirow,lineno,bm,bbm,fancyvrb,ifpdf}
\usepackage{hyperref}
\usepackage[utf8]{inputenc}
\allowdisplaybreaks

\pdfoutput=1

\newcommand{\GeV}      {~\mathrm{GeV}}
\newcommand{\TeV}      {~\mathrm{TeV}}
\newcommand{\MeV}      {~\mathrm{MeV}}

\newcommand{\fb}      {~\mathrm{fb}}
\newcommand{\ab}      {~\mathrm{ab}}

\def\beqn{\begin{eqnarray}}
\def\eeqn{\end{eqnarray}}
\def\beqs{\begin{subequations}}
\def\eeqs{\end{subequations}}
\def\beq{\begin{equation}}
\def\eeq{\end{equation}}
\def\ba{\begin{array}}
\def\ea{\end{array}}

\def\non{\nonumber\\}
\def\hf{\frac{1}{2}}

\def\mL{\mathcal{L}}

\def\mO{\mathcal{O}}

\begin{document}
\title{\Large Future prospects of mass-degenerate Higgs bosons in the $CP$-conserving two-Higgs-doublet model}
\bigskip

\author{Ligong Bian$^{1\,,2}$}
\email{lgbycl@cqu.edu.cn}
\author{Ning Chen$^{3\,,4}$}
\email{ustc0204.chenning@gmail.com}
\author{Wei Su~$^{5\,,6\,,7}$}
\email{weisv@itp.ac.cn}
\author{Yongcheng Wu~$^{8}$}
\email{ycwu@physics.carleton.ca}
\author{Yu Zhang~$^{4\,,9}$}
\email{dayu@nju.edu.cn}
\affiliation{
	$^1$Department of Physics, Chongqing University, Chongqing 401331, China
	\\
    $^2$ Department of Physics, Chung-Ang University, Seoul 06974, Korea
	\\
	$^3$ School of Physics, Nankai University, Tianjin 300071, China
	\\
	$^4$ CAS Center for Excellence in Particle Physics, Beijing 100049, China
	\\
	$^5$ CAS Key Laboratory of Theoretical Physics, Institute of Theoretical Physics, Chinese Academy of Sciences, Beijing 100190, China
	\\
	$^6$ School of Physics, University of Chinese Academy of Sciences, Beijing 100049, China
	\\
	$^7$ Department of Physics, University of Arizona, Tucson, AZ 85721
	\\
	$^8$ Ottawa-Carleton Institute for Physics, Carleton University, 1125 Colonel By Drive, Ottawa, Ontario K1S 5B6, Canada 
	\\
	$^9$ School of Physics, Nanjing University, Nanjing, Jiangsu, 210093, China
}
\date{\today}

\begin{abstract}
The scenario of two mass-degenerate Higgs bosons within the general two-Higgs-doublet model (2HDM) is revisited. 
We focus on the global picture when two $CP$-even Higgs bosons of $h$ and $H$ are nearly mass-degenerate. 
A global fit to the signal strength of the 125 GeV Higgs measured at the LHC is performed. 
Based on the best-fit result of the 2HDM mixing angles $(\alpha,\beta)$, theoretical constraints, charged and $CP$-odd Higgs boson direct search constraints and the electroweak precision constraints are imposed to the 2HDM parameter space. 
We present the signal predictions of the $(4b\,, 2b\,2\gamma)$ channels for the benchmark models at the LHC 14 TeV runs.
We also study the direct Higgs boson pair productions at the LHC, and the Z-associated Higgs boson pair production search at the ILC 500 GeV runs, as well as the indirect probes at the CEPC 250 GeV run.
We find that the mass-degenerate Higgs boson scenario in the Type-II 2HDM can be fully probed by these future experimental searches.
\end{abstract}

\maketitle
\baselineskip=16pt

\pagenumbering{arabic}

\vspace{1.0cm}
\tableofcontents

\newpage


\section{Introduction}
\label{section:intro}

When the $125\,\GeV$ Higgs boson was discovered at the LHC $7\oplus 8\,\TeV$ runs~\cite{Aad:2012tfa,Chatrchyan:2012xdj}, the experimental measurements of the $\gamma\gamma$ signal rates from both ATLAS and CMS collaborations were both enhanced relative to the standard model (SM) predictions.
It was suggested in Refs.~\cite{Gunion:2012gc, Gunion:2012he, Ferreira:2012nv, Chabab:2014ara} that the observed signals at $\sim 125\,\GeV$ may arise from two mass-degenerate Higgs bosons.~\footnote{The previous estimations of the signal rates were performed by summing up the cross sections times decay branching fractions of individual Higgs boson~\cite{Gunion:2012gc, Gunion:2012he,Craig:2013hca}. 
Recently, it was pointed out in Ref.~\cite{Chen:2016oib} that the quantum interference effect should be taken into account for the signal rates from two $CP$-even Higgs bosons (see also Ref.~\cite{Das:2017tob} for the NMSSM case).
This effect was found to be significant when the mass splitting are comparable or smaller than the total decay widths of two nearly degenerate Higgs bosons.}
Even though the obvious enhancement of the $\gamma\gamma$ rate is not shown at the LHC run-II, the scenario of two mass-degenerate Higgs bosons around 125 GeV still deserves investigation.
After all, it is really challenging to distinguish this possibility from the single Higgs boson case by direct measurements of the Higgs boson mass, given that the energy resolutions of photons and leptons are typically of $\sim\mO(1)\,\GeV$ at the LHC~\cite{Aad:2014nim,Aad:2014rra,Chatrchyan:2012xi,Khachatryan:2015hwa,Khachatryan:2015iwa}.
The current mass uncertainties from the CMS measurements are $\sim 0.24\,\GeV$~\cite{Chatrchyan:2012jja,Khachatryan:2014jba}.
The direct measurements of the Higgs boson(s) at $125\,\GeV$ involve their gauge couplings and Yukawa couplings at the leading order (LO).
Alternatively, one may constrain such a scenario from a global point of view, by imposing various theoretical and experimental constraints.

In this work, we study the future experimental prospects of probing the mass-degenerate $125\,\GeV$ Higgs bosons at both high-luminosity (HL) LHC runs and the future high-energy colliders.
Our discussions are made in the context of the $CP$-conserving (CPC) general 2HDM.
This scenario can be constrained from the current LHC searches for the $CP$-odd Higgs boson $A$ via the $h/H+Z$ decay channel. 
Furthermore, we suggest to distinguish the $h/H$ mass-degenerate case from the single resonance case through the probes of the Higgs boson self couplings.
This can be done by searching for the Higgs boson pair production processes at both LHC and the future high-energy colliders.
For the new physics (NP) models involving a single $125\,\GeV$ Higgs boson, it is quite often that the modified Higgs cubic self couplings (including the additional resonances) are the only sources to modify the Higgs pair production cross sections. 
The signal rates for various final states can be estimated by using the SM-like Higgs boson decay branching fractions (see Ref.~\cite{Liu:2013woa} for the summary).
There have been extensive discussions of the Higgs boson pair productions in various beyond standard model (BSM) NP models~\cite{Bian:2016awe, Cacciapaglia:2017gzh, Grober:2017gut, Chalons:2017wnz, Dawson:2017jja, Basler:2017uxn}.
Currently, two most sensitive search modes for the Higgs boson pair productions at the LHC $13\,\TeV$ run are $(4b\,, 2b\,2\gamma)$~\cite{ATLAS:2016ixk,CMS:2016tlj,CMS:2016rec,CMS:2017orf,TheATLAScollaboration:2016ibb,CMS:2016vpz}.
For the Higgs pair productions with two mass-degenerate Higgs bosons, one may expect: (i) the deviation of Higgs cubic self-couplings from the SM predictions, and (ii) the existence of multiple Higgs cubic self-couplings, hence, multiple processes contributing to each final state.

The layout of this paper is described as follows. 
In Sec.~\ref{section:degenerate}, we review the scenario of degenerate Higgs bosons in the framework of CPC 2HDM.
The LHC measurements of the Higgs signal strengths are used for the global fit, where we simplify the discussion with negligible quantum interference and mixing effects. 
This can be achieved by assuming sufficiently large mass splitting.
Other constraints, such as the perturbative unitarity and stability of the 2HDM potential, the EW precision tests, as well as the LHC direct searches for the $CP$-odd Higgs boson $A$, are also considered for the 2HDM with mass-degenerate $h/H$.
The benchmark points are suggested for both Type-I and Type-II 2HDM.
In Sec.~\ref{section:Hpair_LHC}, we study the gluon-gluon fusion (ggF) productions of Higgs pairs at the LHC for the degenerate Higgs scenario. 
We compare the signal predictions from various final states of $(4b\,, 2b\,2\gamma)$ with the corresponding SM predictions at the LO.
Their cross sections are generally varying with different soft mass terms of $m_{12}$ in the 2HDM potential.
In particular, we find that the signal rates of $(4b\,, 2b\,2\gamma)$ final states are always moderately enhanced with respect to the SM predictions.
The corresponding significances are estimated for the $h/H$ mass-degenerate case as well.
In Sec.~\ref{section:Hpair_ee}, we discuss the capability of distinguishing the $h/H$ mass-degenerate scenario at the future high-energy $e^+ e^-$ colliders.
We show the indication from the precise measurement of cross sections of mass-degenerate Higgs bosons with $Z$-boson for this scenario at the circular electron-positron collider (CEPC). 
Furthermore, the direct production of Higgs boson pairs associated with $Z$-boson at the ILC can probe the $h/H$ mass-degenerate scenario in the Type-II 2HDM.
The summaries are given in Sec.~\ref{section:conclusion}.


\section{The mass-degenerate Higgs bosons in the 2HDM}
\label{section:degenerate}

\subsection{The global fit to the mass-degenerate Higgs boson signals at the LHC}

In the CPC 2HDM, there are five Higgs bosons of $(h\,,H\,,A\,,H^\pm)$ in the scalar mass spectrum.
The review of the 2HDM setup and the related LHC phenomenology can be found in Refs.~\cite{Craig:2013hca,Branco:2011iw}.
The Lagrangian for the general 2HDM is written as follows
\beqs
\beqn
\mL&=&\mL_{\rm kin} + \mL_{\rm Yukawa}-V(\Phi_1\,,\Phi_2)\,,\\[2mm]
\mL_{\rm kin}&=&|D_{\mu}\Phi_1|^2+|D_\mu \Phi_2|^2\,,\label{eq:2HDM_kinematic}\\[2mm]
V(\Phi_1\,,\Phi_2)&=&m_{11}^2|\Phi_1|^2+m_{22}^2|\Phi_2|^2- m_{12}^2 (\Phi_1^\dag\Phi_2+H.c.)\non[2mm]
&&+\hf\lambda_1(\Phi_1^\dag\Phi_1)^2+\hf\lambda_2(\Phi_2^\dag \Phi_2)^2+\lambda_3|\Phi_1|^2 |\Phi_2|^2+\lambda_4 |\Phi_1^\dag \Phi_2|^2\non[2mm]
&&+\hf \lambda_5 \Big[ (\Phi_1^\dag\Phi_2)^2+H.c.   \Big]\,.\label{eq:2HDM_potential}
\eeqn
\eeqs
All parameters are assumed to be real for the CPC case.
Very often, a softly broken $\mathbb{Z}_2$ symmetry, under which two Higgs doublets transform as $(\Phi_1\,,\Phi_2)\to (\Phi_1\,, - \Phi_2)$, is also assumed to eliminate the possible $\lambda_{6\,,7}$ couplings in the 2HDM potential.
One has $m_{12}=0$ when the $\mathbb{Z}_2$ symmetry is exact. 
Two Higgs doublets of $\Phi_{1,\,2}$ can be expressed in terms of components as
\beqn\label{eq:Higgs_doublet}
&&\Phi_1=\left(\begin{array}{c} \pi_1^+ \\  \frac{1}{ \sqrt{2} } ( v_1 + h_1 + i\pi_1^0 )  \end{array} \right)\,,\qquad \Phi_2=\left(\begin{array}{c} \pi_2^+ \\  \frac{1}{ \sqrt{2} } ( v_2 + h_2 + i\pi_2^0 )  \end{array} \right)\,,
\eeqn
with two Higgs vacuum expectation values (VEVs) and their ratios being
\beqn\label{eq:Higgs_vev}
&&v_1^2+v_2^2=(\sqrt{2}G_F)^{-1}\simeq(246\,\GeV)^2 \,,\qquad t_\beta \equiv v_2/v_1 \,.
\eeqn
Here, $\pi_{1\,,2}^0$ are pseudoreal components, whose linear combinations of $A= - s_\beta \pi_1^0 + c_\beta \pi_2^0$ and $G=c_\beta \pi_1^0 + s_\beta \pi_2^0$ are $CP$-odd Higgs boson and neutral Nambu-Goldstone boson, respectively.
$\pi_{1\,,2}^+$ (and their complex conjugates) are complex scalar fields, whose linear combinations of $H^\pm= - s_\beta \pi_1^\pm + c_\beta \pi_2^\pm$ and $G^\pm=c_\beta \pi_1^\pm + s_\beta \pi_2^\pm$ are charged Higgs bosons and charged Nambu-Goldstone bosons, respectively.

For further discussion, we list the dimensionless Higgs gauge couplings and Yukawa couplings as follows
\beqn
\mL&\supset& \sum_{h_i=h\,,H} \Big[ - \frac{m_f}{v}  \xi_i^f \bar f f + a_i \left( 2  \frac{m_W^2}{v} W_\mu^+ W^{-\,\mu} +  \frac{m_Z^2}{v} Z_\mu Z^\mu \right)  \Big] h_i\non 
&-& \frac{m_f}{v} \xi_A^f \bar f i \gamma_5 f A \,,
\eeqn
with
\beqs\label{eqs:Higgs_coup}
\beqn
\textrm{Type-I}&:& \xi_h^f= s_{\beta-\alpha} + \frac{c_{\beta-\alpha}}{t_\beta} \,,\qquad \xi_{H}^f=c_{\beta-\alpha} - \frac{s_{\beta-\alpha}}{t_\beta}\non
&& \xi_A^u= \frac{1}{t_\beta}\,, \qquad \xi_A^{d\,, \ell}= - \frac{1}{t_\beta}\,,\\
\textrm{Type-II}&:&\xi_h^u=s_{\beta-\alpha} + \frac{c_{\beta-\alpha}}{t_\beta}\,,\qquad \xi_h^{d\,,\ell}= s_{\beta-\alpha}-c_{\beta-\alpha}t_\beta \,,\non
&&\xi_H^u=c_{\beta-\alpha}- \frac{s_{\beta-\alpha}}{t_\beta}\,,\qquad \xi_H^{d\,,\ell}=c_{\beta-\alpha}+ s_{\beta-\alpha} t_\beta\,,\non
&& \xi_A^u= \frac{1}{t_\beta}\,,\qquad \xi_A^{d\,,\ell} = t_\beta\,,\\
&& a_h= s_{\beta-\alpha}\,,\qquad a_H = c_{\beta-\alpha}\,.
\eeqn
\eeqs
Here, $\alpha$ represents the mixing angle between two $CP$-even Higgs bosons of $(h\,,H)$.

\begin{table}[htbp]
\vspace{0.2cm}
\begin{center}
\begin{tabular}{c|c|c|c|c|c}
\hline\hline
Decays & Productions & ATLAS & Ref & CMS & Ref   \\\hline\hline
$\gamma\gamma$ & ggF  & $1.32\pm 0.38$   &\cite{Aad:2014eha}  & $1.12^{+0.37}_{-0.32}$   &\cite{Khachatryan:2014ira}           \\\hline
$\gamma\gamma$ & VBF    & $0.8\pm 0.7$  & \cite{Aad:2014eha}  & $1.58^{+0.77}_{-0.68}$ & \cite{Khachatryan:2014ira}             \\\hline
$\gamma\gamma$ & WH     & $1.0\pm 1.6$    & \cite{Aad:2014eha}           & $\cdots$           & $\cdots$          \\\hline
$\gamma\gamma$ & ZH      & $0.1^{+3.7}_{-0.1}$              & \cite{Aad:2014eha}           & $\cdots$             & $\cdots$      \\\hline
$\gamma\gamma$ & VH      & $\cdots$        & $\cdots$   & -$0.16^{+1.16}_{-0.79}$   & \cite{Khachatryan:2014ira}            \\\hline
$\gamma\gamma$ & ttH      & $1.6^{+2.7}_{-1.8}$   & \cite{Aad:2014eha}   & $2.69^{+2.51}_{-1.81}$          & \cite{Khachatryan:2014ira}           \\\hline\hline
$ZZ$ & ggF,ttH,bbH      & $1.7^{+0.5}_{-0.4}$     & \cite{Aad:2014eva}            & $\cdots$                  &  $\cdots$                 \\\hline
$ZZ$ & ggF,ttH                     & $\cdots$                                           & $\cdots$                                          & $0.80^{+0.46}_{-0.36}$            &\cite{Chatrchyan:2013mxa}             \\\hline
$ZZ$ & VBF,VH                    & $0.3^{+1.6}_{-0.9}$             &\cite{Aad:2014eva}             & $1.7^{+2.2}_{-2.1}$                     &\cite{Chatrchyan:2013mxa}              \\\hline\hline
$W^+ W^-$ & ggF                        & $1.02^{+0.29}_{-0.26}$       &\cite{ATLAS:2014aga}         & $0.74^{+0.22}_{-0.20}$          &\cite{Chatrchyan:2013iaa}            \\\hline
$W^+ W^-$ & VBF                       & $1.27^{+0.53}_{-0.45}$       & \cite{ATLAS:2014aga}        & $0.60^{+0.57}_{-0.46}$         &\cite{Chatrchyan:2013iaa}           \\\hline
$W^+ W^-$ & VH                         & $\cdots$                                            & $\cdots$                                           & $0.39^{+1.97}_{-1.87}$          & \cite{Chatrchyan:2013iaa}           \\\hline\hline
$b \bar b$ & ttH                            & $1.5\pm 1.1$                        &\cite{Aad:2015gra}              & $1.2^{+1.6}_{-1.5}$                    & \cite{Khachatryan:2015ila}           \\\hline
$b \bar b$ & VH                           & $0.51^{+0.40}_{-0.37}$         & \cite{Aad:2014xzb}            & $1.0\pm 0.5$                               &\cite{Chatrchyan:2013zna}          \\\hline\hline
$\tau^+ \tau^-$ & ggF                  & $2.0^{+1.5}_{-1.2}$               & \cite{Aad:2015vsa}            & $1.07\pm 0.46$                    & \cite{Chatrchyan:2014nva}           \\\hline
$\tau^+ \tau^-$ & VBF,VH           & $1.24^{+0.59}_{-0.54}$          &\cite{Aad:2015vsa}             & $\cdots$                                             & $\cdots$                                                      \\\hline
$\tau^+ \tau^-$ & VBF                 & $\cdots$                                              & $\cdots$                                          & $0.94 \pm 0.41$                   &\cite{Chatrchyan:2014nva}            \\\hline
$\tau^+ \tau^-$ & VH                  & $\cdots$                                               & $\cdots$                                          & -$0.33 \pm 1.02$                    &\cite{Chatrchyan:2014nva}           \\\hline\hline
\end{tabular}
\caption{Signal strengths of Higgs searches measured by the ATLAS and  CMS collaborations, 
 for various decay and production channels for the $\sqrt{s}=7 \oplus 8\,\TeV$ runs.
}
\label{tab:LHCRunI_Higgs}
\end{center}
\end{table}

\begin{table}[htbp]
\vspace{0.2cm}
\begin{center}
\begin{tabular}{c|c|c|c|c|c}
\hline\hline
Decays & Productions & ATLAS & Ref & CMS & Ref   \\\hline\hline
$\gamma\gamma$ & ggF     & $0.80_{-0.18}^{+0.19}$  &  \cite{ATLAS:2017myr, ATLAS:2017ovn}  & $1.11_{-0.18}^{+0.19}$  &\cite{CMS:2017jkd}    \\\hline
$\gamma\gamma$ & VBF & $2.1\pm 0.6$  & \cite{ATLAS:2017myr, ATLAS:2017ovn}  & $0.5_{-0.5}^{+0.6}$   & \cite{CMS:2017jkd}   \\\hline
$\gamma\gamma$ & VH  & $0.7_{-0.8}^{+0.9}$  & \cite{ATLAS:2017myr, ATLAS:2017ovn}  & $2.3_{-1.0}^{+1.1}$    & \cite{CMS:2017jkd}  \\\hline
$\gamma\gamma$ & ttH      & $0.5\pm 0.6$  & \cite{ATLAS:2017myr, ATLAS:2017ovn}   & $2.2_{-0.8}^{+0.9}$  & \cite{CMS:2017jkd}   \\\hline\hline
$ZZ$ & ggF    & $1.11_{-0.22}^{+0.25}$   & \cite{ATLAS:2017ovn, ATLAS:2017cju}    &    $1.20_{-0.21}^{+0.22}$   &   \cite{Sirunyan:2017exp}     \\\hline
$ZZ$ & VBF    & $4.0_{-1.5}^{+1.8}$   &  \cite{ATLAS:2017ovn, ATLAS:2017cju}  & $0.05_{-0.05}^{+1.03}$            &\cite{Sirunyan:2017exp}   \\\hline
$ZZ$ &  VH     & $0\pm 1.9$  &\cite{ATLAS:2017ovn, ATLAS:2017cju}  & $0\pm 2.83\,,\textrm{or } 0\pm 2.66$    &\cite{Sirunyan:2017exp}   \\\hline
$ZZ$ &  ttH     & $0\pm 3.9$    &\cite{ATLAS:2017ovn, ATLAS:2017cju}  & $0\pm 1.19$     &\cite{Sirunyan:2017exp}      \\\hline\hline
$W^+ W^-$ & ggF      & $\cdots$      & $\cdots$        & $1.02\pm 0.27$         &\cite{CMS-PAS-HIG-16-021}           \\\hline
$W^+ W^-$ & VBF  & $1.7_{-0.9}^{+1.2}$  & \cite{ATLAS-CONF-2016-112}      & $\cdots$        &$\cdots$           \\\hline
$W^+ W^-$ & WH   & $3.2_{-4.2}^{+4.4}$  & \cite{ATLAS-CONF-2016-112}     & $\cdots$          & $\cdots$        \\\hline\hline
$W^+ W^-$ & VBF+VH       &$\cdots$      &$\cdots$                 & $0.89\pm 0.67$          & \cite{CMS-PAS-HIG-16-021}        \\\hline\hline
$b \bar b$ & VH          & $1.20_{-0.36}^{+0.42}$      & \cite{Aaboud:2017xsd}            & $\cdots$               & $\cdots$          \\\hline\hline
$\tau^+ \tau^-$ & ggF           &$\cdots$         &$\cdots$      & $0.84\pm 0.89$                    & \cite{Sirunyan:2017khh}           \\\hline
$\tau^+ \tau^-$ & VBF            &$\cdots$             &$\cdots$      &   $1.11_{-0.35}^{+0.34}$  &\cite{Sirunyan:2017khh}            \\\hline
$\tau^+ \tau^-$ & ttH              &$\cdots$             &$\cdots$    & $0.72_{-0.53}^{+0.62}$    &\cite{CMS:2017lgc}           \\\hline\hline
\end{tabular}
\caption{Signal strengths of Higgs searches measured by the ATLAS and CMS collaborations, for various decay and production channels for the $\sqrt{s}=13 \,\TeV$ runs.
}
\label{tab:LHCRunII_Higgs}
\end{center}
\end{table}

\begin{figure}[htb]
\centering
\includegraphics[width=0.45\textwidth]{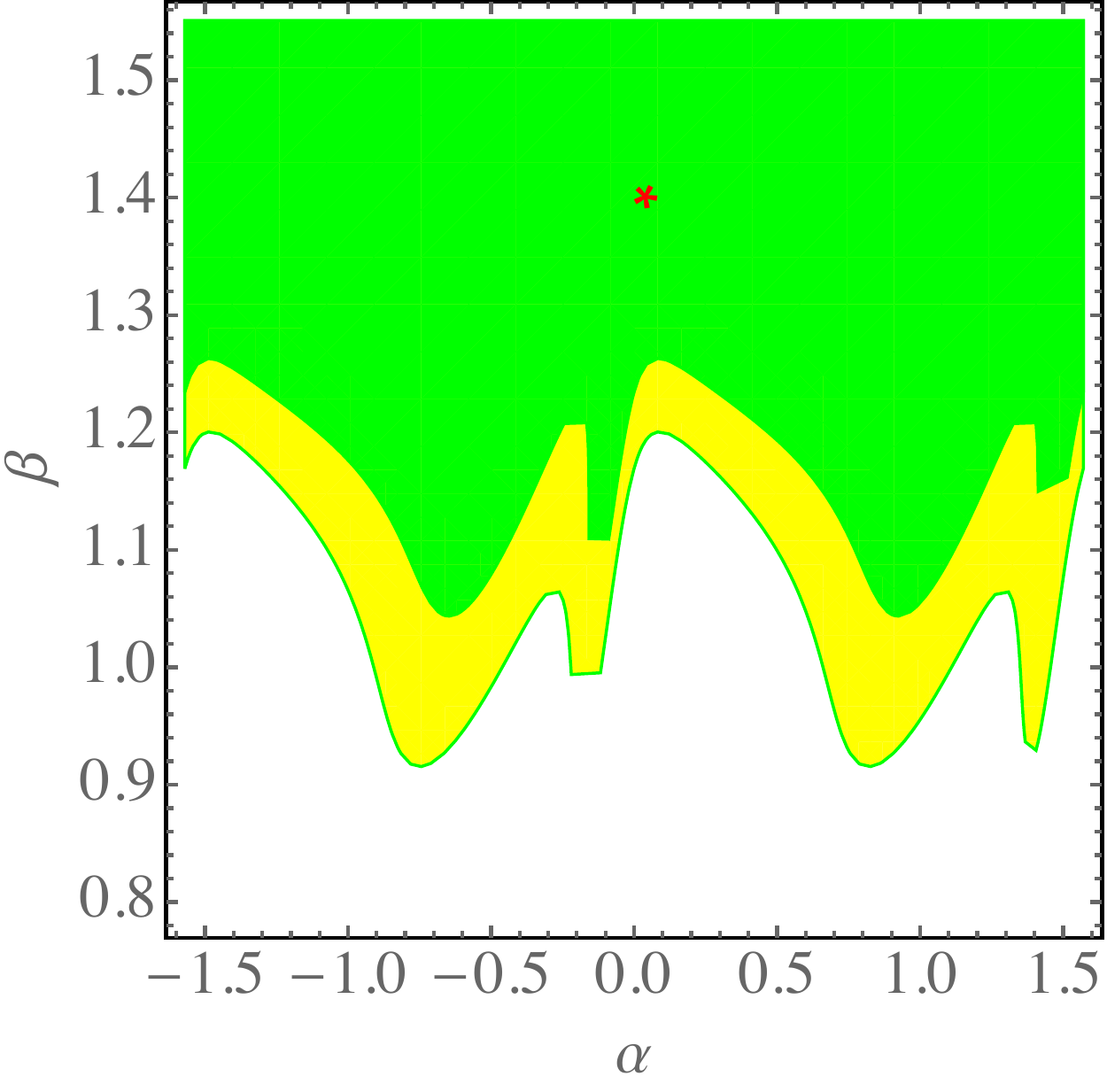}
\includegraphics[width=0.45\textwidth]{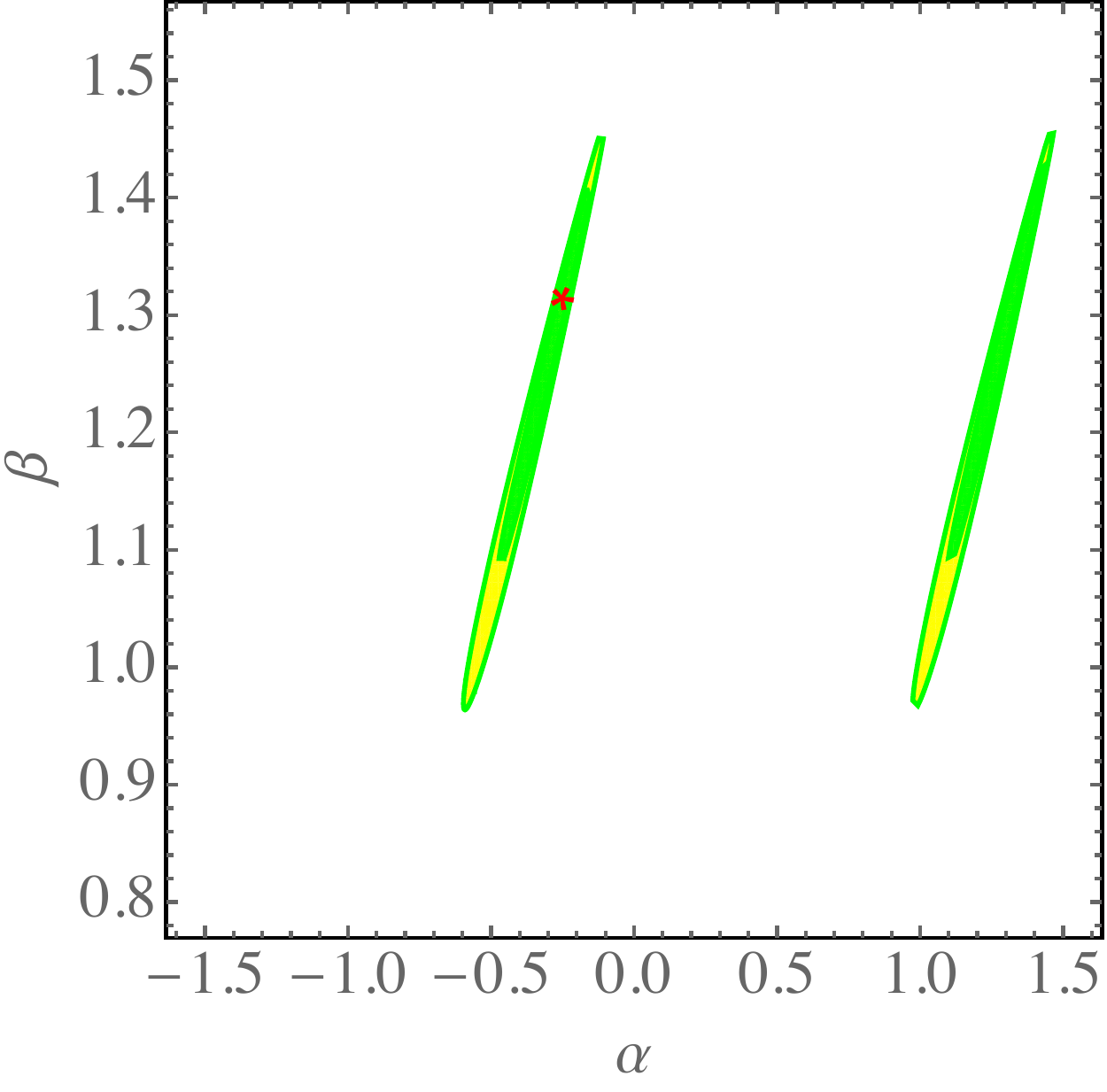}
\caption{\label{fig:hH_fit}
The global fit of two mass-degenerate Higgs bosons in the Type-I (left panel) and Type-II (right panel) 2HDM on the $(\alpha\,,\beta)$ plane.
The yellow and green regions are the $(1\,,2)\,\sigma$ allowed regions of the LHC $7\oplus 8\oplus 13\,\TeV$ Higgs data fitting, and the benchmark points are marked by red stars.
}
\end{figure}

\begin{table}[htb]
\begin{center}
\resizebox{\textwidth}{!}{
\begin{tabular}{c|c|c}
\hline\hline
 $M_h\approx M_H$ & Type-I & Type-II   \\\hline
 $(\alpha\,,\beta\,,\Gamma_{\rm tot})$  &  $(0.04263\,, 1.3995\,,4.16\,\MeV)$  & $(-0.2495\,, 1.3121\,, 41.38\,\MeV)$  \\ \hline
 $(\Gamma_h\,, \Gamma_H)$  &  $(4.11\,\MeV\,, 0.05\,\MeV)$  & $(3.89\,\MeV\,, 37.49\,\MeV)$  \\ \hline
$({\rm Br}[h\to b \bar b]\,, {\rm Br}[H\to b \bar b])$ & $(58.80\,\%\,, 8.68\,\%)$  & $(56.23\,\%\,, 89.85\,\%)$   \\ \hline 
$({\rm Br}[h\to \tau^+ \tau^-]\,, {\rm Br}[H\to \tau^+ \tau^-])$ & $(6.44 \,\%\,, 0.95 \,\%)$  & $(6.16 \,\%\,, 9.84 \,\%)$   \\ \hline 
$({\rm Br}[h\to W^+ W^-]\,, {\rm Br}[H\to W^+ W^-])$ & $(20.35 \,\%\,, 77.86\,\% )$  & $( 22.50\,\%\,,- )$   \\ \hline 
$({\rm Br}[h\to ZZ]\,, {\rm Br}[H\to ZZ])$ & $( 2.50\,\%\,,9.56\,\%)$  & $(2.76 \,\%\,, - )$   \\ \hline 
$({\rm Br}[h\to \gamma\gamma]\,, {\rm Br}[H\to \gamma\gamma])$ & $( 0.21\,\%\,, 1.22\,\%)$  & $( 0.24\,\%\,, -)$   \\ \hline 
$({\rm Br}[h\to gg]\,, {\rm Br}[H\to gg])$ & $( 8.73\,\%\,, 1.29\,\%)$  & $( 9.04\,\%\,, 0.28\,\%)$   \\ \hline \hline
\end{tabular}
}
\caption{
The best-fit points of $(\alpha\,,\beta)$ for the mass-degenerate Higgs bosons of $M_h\approx M_H= 125\,\GeV$ in both Type-I and Type-II 2HDM.
The decay widths of $(\Gamma_h\,, \Gamma_H)$, and decay branching ratios are listed, where the decay branching ratios smaller than $10^{-4}$ are neglected.}
\label{tab:BM_hH}
\end{center}
\end{table}

The overall signal rates are controlled by the input parameters of $(\alpha\,,\beta)$ in the CPC 2HDM, which is manifest from the couplings in Eqs.~\eqref{eqs:Higgs_coup}.
A global fit to the $h/H$ degenerate scenario respect to $(\alpha\,,\beta)$ is thus performed, and this is done by the $\chi^2$ fit to the LHC data defined as
\beqn\label{eq:chi2}
\chi^2&=& \sum_{\rm PD} \Big( \frac{\mu_{\rm th}^{\rm PD} - \mu_{\rm exp}^{\rm PD}}{ \sigma_{\rm exp}^{\rm PD} }  \Big)^2\,,
\eeqn
where the current LHC measurements of signal strengths and errors of $(\mu_{\rm exp}^{\rm PD}\,, \sigma_{\rm exp}^{\rm PD})$ are summarized in Tables~\ref{tab:LHCRunI_Higgs} and~\ref{tab:LHCRunII_Higgs} for the run-I and run-II data, respectively.

This scenario was previously explored in Refs.~\cite{Gunion:2012gc, Gunion:2012he, Ferreira:2012nv, Chabab:2014ara, Craig:2013hca}, where the total signal rates for $125\,\GeV$ Higgs boson were estimated by simple summation of $\sigma\times {\rm Br}$ from the individual contribution of $h$ and $H$ as
\beqn\label{eq:hH_sum}
\mu[XX\to h/H\to YY]&=& \frac{|\kappa_{hXX} \kappa_{hYY} |^2}{\Gamma_h/ \Gamma_h^{\rm SM}} + \frac{|\kappa_{HXX} \kappa_{HYY} |^2}{\Gamma_H/ \Gamma_h^{\rm SM}}\,.
\eeqn
This is valid when the mass splitting between two resonances of $h$ and $H$ are sufficiently large such that the quantum interference between two amplitudes are negligible.
Quantitatively, the sufficiently large mass splitting refers to the case when $\Delta M\equiv M_H - M_h \gg \Gamma_H + \Gamma_h$~\cite{Chen:2016oib}.
To simplify our discussions, we still use the simple summation method in Eq.~\eqref{eq:hH_sum} to estimate the total signal rates for various channels. 
The global fit results on the $(\alpha\,,\beta)$ plane are shown in Fig.~\ref{fig:hH_fit}.
The corresponding benchmark points for the mass-degenerate $M_h\approx M_H =125\,\GeV$ cases are listed in Table.~\ref{tab:BM_hH} for both Type-I and Type-II 2HDM.
We also observe that a shift of $\alpha \to \alpha-\pi/2$ leads to equally minimal $\chi^2$ values in both Type-I and Type-II cases.
This corresponds to an interchange between $h$ and $H$ in the mass-degenerate scenario.
Besides the 2HDM parameters of $(\alpha\,, \beta)$, the total decay widths of $\Gamma_h + \Gamma_H$ and the main decay branching fractions are also listed for the mass-degenerate case.
Since the total decay widths of $\Gamma_h+\Gamma_H$ are smaller than $\sim \mO(0.1)\,\GeV$, our simplification in Eq.~\eqref{eq:hH_sum} is valid.
The alignment parameters are $c_{\beta-\alpha}=0.21$ for the Type-I 2HDM and $c_{\beta-\alpha}=0.01$ for the Type-II 2HDM, respectively.
A sizable deviation from the alignment limit is observed in the Type-I benchmark point.
For the Type-II case, meanwhile, the $H$ is gauge-phobic and decays mostly into fermionic final states of $(b\bar b\,, \tau^+ \tau^-)$.
Throughout the context below, we shall always use the best-fit points of $(\alpha\,,\beta)$ in Table.~\ref{tab:BM_hH} for the phenomenology discussions in the mass-degenerate scenario.

Besides the best-fit points from the current LHC Higgs data, we shall further impose theoretical and experimental constraints to the 2HDM mass spectrum of $(M_A\,, M_\pm\,, m_{12})$ in the following context, for the mass-degenerate Higgs boson scenario.
We shall show that mass-degenerate Higgs boson scenario has allowed 2HDM parameter space with all these constraints imposed.

\subsection{The charged Higgs boson and EW precision constraints to the 2HDM}

It is known that the charged Higgs bosons of $H^\pm$ contribute to the flavor-changing neutral current (FCNC) rare decay processes, such as $b\to s\gamma$ transition.
The latest measurement is from the Belle Collaboration~\cite{Belle:2016ufb}, and the implication to the CPC 2HDM was carried out in Refs.~\cite{Misiak:2015xwa,Misiak:2017bgg,Arbey:2017gmh,Arnan:2017lxi}.
By imposing the FCNC constraints to the benchmark models in Table.~\ref{tab:BM_hH}, we get $M_\pm\gtrsim 590\,\GeV$ in the Type-II 2HDM, while the lower mass bound in the Type-I 2HDM is negligible, as compared to the direct collider constraints.
Besides, the direct searches for the charged Higgs bosons at the LHC were performed in Refs.~\cite{Khachatryan:2015qxa,Aad:2015typ,Aaboud:2016dig}.
Here, we shall only consider the FCNC constraints to the charged Higgs boson mass, and leave the direct LHC search limits to the charged Higgs bosons in the context of the mass-degenerate Higgs bosons.
This is valid because: (i) the FCNC constraints are only relevant to the charged Higgs Yukawa couplings, and (ii) the decay modes of the charged Higgs bosons can be significantly modified in the mass-degenerate Higgs boson scenario.
To simplify, we shall take $M_\pm=M_A$ for the Type-I 2HDM~\footnote{As we shall see below, the specific mass ranges of $(M_\pm\,, M_A)$ do not play a role in the Higgs boson pair productions at the LHC or ILC. Without loss of generality, we make such simplification of $M_\pm=M_A$. }, and fix $M_\pm=600\,\GeV$ for the Type-II 2HDM below~\footnote{When taking the unitarity bound into account, it turns out that the charged Higgs boson mass cannot exceed $\sim 625\,\GeV$. Thus, a fixed $M_\pm=600\,\GeV$ is taken to compromise the joint constraints from the FCNC rare decay and unitarity for the Type-II case. }.

We consider the constraints from the EW precision tests~\cite{He:2001tp,Grimus:2008nb,Haber:2010bw} to the 2HDM with mass-degenerate $h/H$.
The most general expressions for $(\Delta S\,, \Delta T)$ in the CPC 2HDM~\cite{He:2001tp} read
\beqs\label{eqs:2HDM_EWPD}
\beqn
\Delta\,S&=&\frac{1}{\pi\, m_Z^2} \Big\{  \Big[ {\cal B}_{22}( m_Z^2\,; M_H^2\,, M_A^2 ) - {\cal B}_{22}( m_Z^2\,; M_\pm^2\,, M_\pm^2)  \Big] \non
&&+ \Big[ {\cal B}_{22}( m_Z^2\,; M_h^2\,, M_A^2 )- {\cal B}_{22}( m_Z^2\,; M_H^2\,, M_A^2 ) + {\cal B}_{22}( m_Z^2\,; m_Z^2\,, M_H^2 ) - {\cal B}_{22}( m_Z^2\,; m_Z^2\,, M_h^2 ) \non
&& - m_Z^2 {\cal B}_0 ( m_Z\,; m_Z\,, M_H^2 ) + m_Z^2 {\cal B}_0 ( m_Z\,; m_Z\,, M_h^2 )    \Big] c_{\beta-\alpha}^2  \Big\}\,,\\
\Delta T&=& \frac{1}{ 16\pi\, m_W^2\, s_W^2 } \Big\{  \Big[ F( M_\pm^2 \,, M_A^2 ) +    F( M_\pm^2\,, M_H^2) - F(M_A^2\,, M_H^2)  \Big]\non
&&+   \Big[  F( M_\pm^2 \,, M_h^2 ) -  F( M_\pm^2\,, M_H^2) - F( M_A^2\,, M_h^2 )+ F(M_A^2\,, M_H^2)   \non
&& + F( m_W^2\,, M_H^2) - F(m_W^2\,, M_h^2 ) - F(m_Z^2 \,, M_H^2 ) + F(m_Z^2\,, M_h^2)\non
&& + 4 m_Z^2 \overline B_0 (m_Z^2\,, M_H^2\,, M_h^2)   - 4 m_W^2 \overline B_0 ( m_W^2\,, M_H^2\,, M_h^2 )  \Big] c_{\beta-\alpha}^2 \Big\}\,,
\eeqn
\eeqs
where we explicitly split these expressions into terms independent of or dependent on the alignment parameter of $c_{\beta-\alpha}$.
The relevant auxiliary functions read
\beqs\label{eqs:STU_aux}
\beqn
F(x_1\,, x_2)&\equiv& \left\{\begin{array}{ll} \frac{x_1 + x_2}{2} - \frac{x_1 x_2}{ x_1- x_2} \ln \frac{x_1}{x_2}  & \hspace{1cm} x_1 \neq  x_2 \vspace{0.2cm} \\
0 & \hspace{1cm} x_1 = x_2  \end{array} \right. \label{eq:STU_F}\\
f(x_1\,, x_2)&\equiv&\left\{\begin{array}{ll} - 2 \sqrt{\Delta} \Big[ \tan^{-1} \frac{x_1 - x_2 +1}{\sqrt{\Delta}} -  \tan^{-1} \frac{x_1 - x_2 -1}{\sqrt{\Delta}}  \Big] & \hspace{1cm} \Delta>0 \vspace{0.2cm} \\
0 & \hspace{1cm} \Delta =0 \vspace{0.2cm} \\
 \sqrt{- \Delta}  \ln \frac{ x_1 + x_2 -1+ \sqrt{- \Delta} }{  x_1 + x_2 -1 - \sqrt{- \Delta} } & \hspace{1cm} \Delta <0 \vspace{0.2cm}  \end{array} \right. \\
 \Delta &=& 2 (x_1 + x_2) - ( x_1 - x_2 )^2 -1 \,,\\
{\cal B}_0 (q^2\,; m_1^2\,, m_2^2)&\equiv& 1 + \frac{1}{2} \Big[ \frac{x_1 + x_2}{ x_1 - x_2 } - (x_1 - x_2)   \Big]\, \ln \frac{x_1}{x_2} + \frac{1}{2} f(x_1\,, x_2) \,,\\
{\cal B}_{22} (q^2\,; m_1^2\,, m_2^2) &\equiv& \frac{q^2}{24} \Big\{  2 \ln q^2 + \ln (x_1 x_2) + \Big[  (x_1 - x_2)^3 - 3 (x_1^2 - x_2^2) + 3(x_1 - x_2)  \Big] \ln \frac{x_1}{x_2} \non
&&- \Big[  2(x_1 - x_2)^2 - 8 (x_1 + x_2) + \frac{10}{3} \Big]\non
&&- \Big[ (x_1 - x_2)^2 - 2 (x_1 + x_2) +1   \Big] f(x_1\,, x_2) - 6 F(x_1\,, x_2)   \Big\} \,,  \label{eq:STU_B22} \\
\overline B_0 (m_1^2\,, m_2^2\,, m_3^2)& \equiv & \frac{m_1^2\ln m_1^2  - m_3^2 \ln m_3^2 }{ m_1^2 - m_3^2  } - \frac{m_1^2\ln m_1^2  - m_2^2 \ln m_2^2 }{ m_1^2 - m_2^2  }  \,,
\eeqn
\eeqs
with $x_i = m_i^2/q^2$.
The current Gfitter fit~\cite{Baak:2014ora} to the EW data gives
\beqn
&&S=0.05\pm 0.11\,,\qquad  T=0.09\pm 0.13\,.
\eeqn
For benchmark models in both Type-I and Type-II cases, the alignment parameters were found to be small as from Table.~\ref{tab:BM_hH}.
Thus, the 2HDM contributions to the $(\Delta S\,, \Delta T)$ are mainly controlled by leading terms in the first lines of Eqs.~\eqref{eqs:2HDM_EWPD}.
By using the definitions of auxiliary functions of \eqref{eq:STU_F} and \eqref{eq:STU_B22}, the $(\Delta S\,, \Delta T)$ can be suppressed with degenerate mass inputs of $M_A=M_\pm$.
Indeed, by using the best-fit $(\alpha\,,\beta)$ inputs for the $h/H$ degenerate cases in Table.~\ref{tab:BM_hH}, we find $(\Delta S\,, \Delta T)\sim (-10^{-4}\,, -10^{-7})$ with $M_A=M_\pm$ for the Type-I 2HDM, and $(\Delta S\,, \Delta T)\sim ( -10^{-4}\,, 10^{-2})$ with $M_A\in (200\,, 300)\,\GeV$ and fixed $M_\pm=600\,\GeV$ input for the Type-II 2HDM~\footnote{Here, we take the mass range of $M_A\in (200\,, 300)\,\GeV$ by considering the perturbative unitarity constraint and the direct search limit of $A\to hZ$ at the LHC.}.

\subsection{The perturbative unitarity and stability constraints to the 2HDM potential}
\label{sec:theoreticalConstraints}

The joint constraints of the perturbative unitarity and tree-level stability conditions to the 2HDM potential turns out to be powerful to bound the heavy scalar masses. 
The conditions to be satisfied for the unitarity constraints to the 2HDM potential are that the absolute values of the following linear combinations of the quartic scalar couplings~\cite{Arhrib:2000is, Kanemura:2015ska}:
\beqn\label{eq:unitarity}
&&a_\pm= \frac{3}{2}(\lambda_1 + \lambda_2 ) \pm \hf \sqrt{ 9(\lambda_1 - \lambda_2)^2 + (2 \lambda_3 + \lambda_4)^2  }\,,\non
&&b_\pm=  \hf \Big[ (\lambda_1 + \lambda_2 ) \pm \sqrt{ (\lambda_1 - \lambda_2)^2 + 4 \lambda_4^2 }   \Big]\,,\non
&&c_\pm=\hf \Big[ (\lambda_1 + \lambda_2 ) \pm \sqrt{ (\lambda_1 - \lambda_2)^2 + 4 \lambda_5^2 }   \Big]\,,\non
&& f_+ = \lambda_3 + 2\lambda_4 + 3\lambda_5\,,\qquad f_-= \lambda_3 + \lambda_5 \,,\qquad f_1=f_2 = \lambda_3 + \lambda_4\,,\non
&& e_1 = \lambda_3 + 2 \lambda_4 - 3\lambda_4\,,\qquad e_2= 2\lambda_3 - \lambda_5\,,\qquad p_1= \lambda_3- \lambda_4\,,
\eeqn
should be smaller than or equal to $8\pi$.
The tree-level vacuum stability conditions for the general 2HDM potential come from the requirement that the scalar potential being bounds from below, which read~\cite{Barroso:2013awa}~\footnote{Recently, it was suggested in Ref.~\cite{Xu:2017vpq} to apply the global minimum condition to constrain the 2HDM potential.
In Ref.~\cite{Staub:2017ktc}, the loop effects to the vacuum stability conditions were found to alleviate the tree-level conditions.}
\beqn\label{eq:stability}
&& \lambda_{1\,,2} \geq0\,,\qquad \lambda_3 \ge - \sqrt{\lambda_1 \lambda_2}\,,\qquad \lambda_3+ \lambda_4 - |\lambda_5| \geq - \sqrt{\lambda_1 \lambda_2}\,.
\eeqn
The quartic self couplings of $\lambda_i$ are related to the Higgs boson masses, the mixing angles, and the soft mass term as follows
\beqs\label{eqs:lambda}
\beqn
\lambda_1&=&\frac{M_h^2 s_\alpha^2+M_H^2 c_\alpha^2 - m_{12}^2 t_\beta }{v^2 c_\beta^2}\,,\label{eq:lam1}\\
\lambda_2&=&\frac{M_h^2 c_\alpha^2+M_H^2 s_\alpha^2 - m_{12}^2 /t_\beta }{v^2 s_\beta^2}\,, \label{eq:lam2} \\
\lambda_3&=&\frac{1}{v^2} \Big[   \frac{(M_H^2-M_h^2)s_\alpha c_\alpha}{s_\beta c_\beta} +  2M_\pm^2 -  \frac{m_{12}^2}{s_\beta c_\beta} \Big]\,,\\
\lambda_4&=&\frac{1}{v^2} ( M_A^2  - 2 M_\pm^2  + \frac{m_{12}^2}{s_\beta c_\beta }  ) \,,\\
\lambda_5&=& \frac{ 1 }{v^2}( \frac{m_{12}^2}{s_\beta c_\beta } - M_A^2)\,.
\eeqn
\eeqs

\begin{figure}[htb]
\centering
\includegraphics[width=0.45\textwidth]{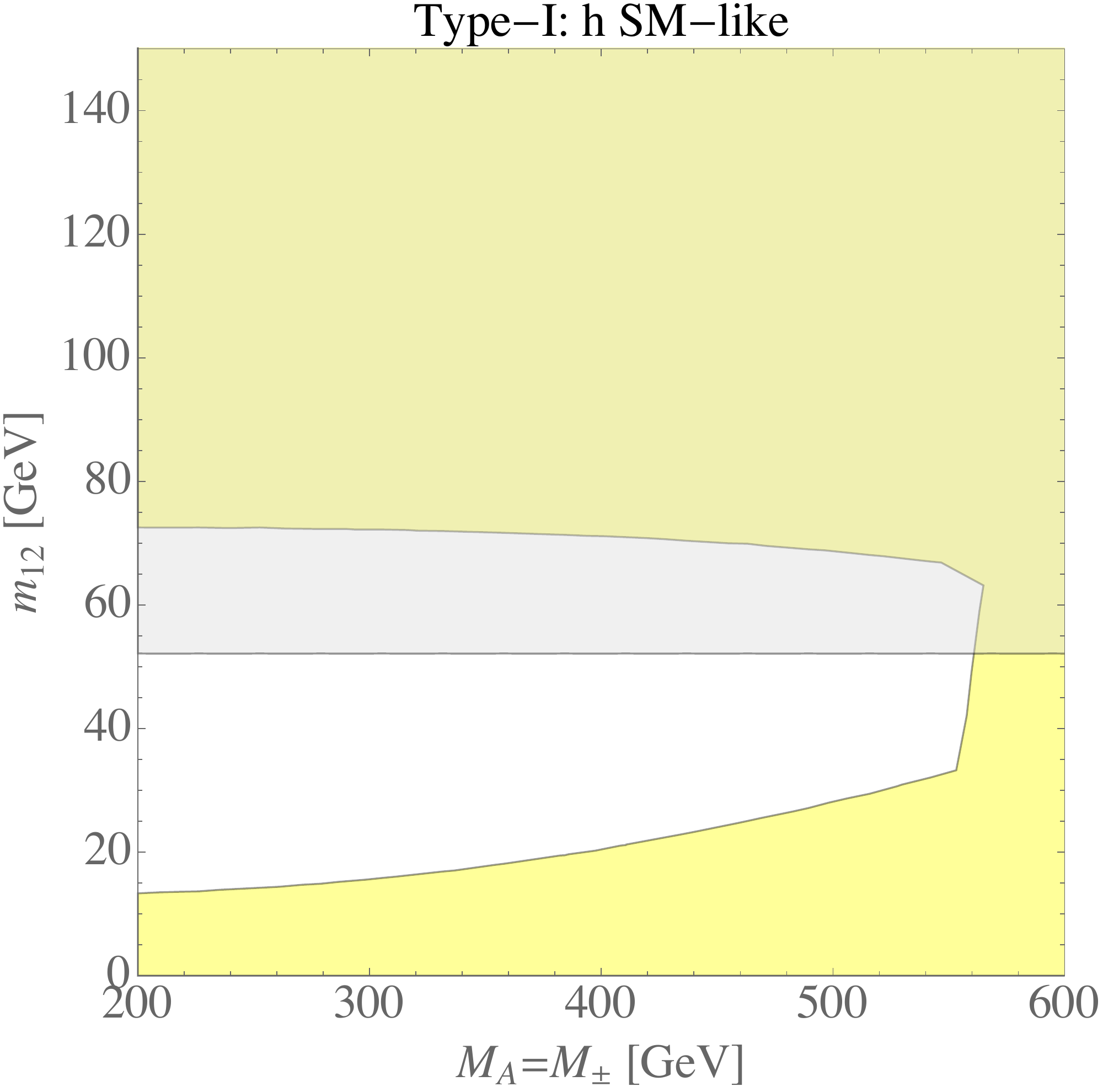}
\includegraphics[width=0.45\textwidth]{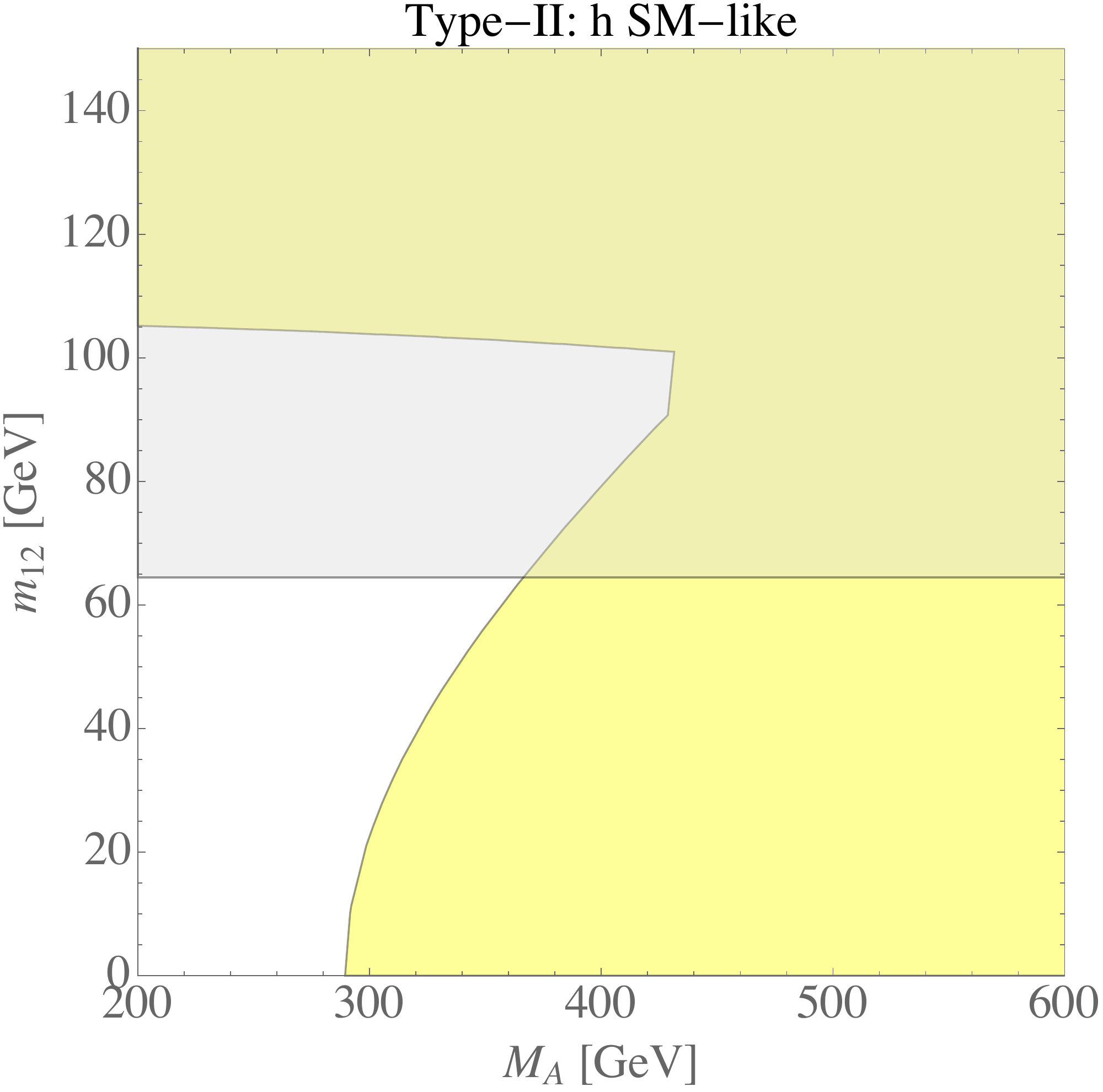}
\caption
{\label{fig:UnitarityStability}
The unitarity (yellow) and stability (gray) excluded regions for Type-I (left) and Type-II (right) 2HDM, with the $M_h\approx M_H=125\,\GeV$ scenario. 
The best-fit points of $(\alpha\,,\beta)=(0.04\,,1.40)$ for Type-I case and $(\alpha\,,\beta)=(-0.25\,, 1.31)$ for Type-II case are taken as in Table.~\ref{tab:BM_hH}.
On the right panel, we fix $M_{\pm}=600$ GeV to evade the B-physics constraints in the Type-II case. 
}
\end{figure}

By combining the constraints in Eq.~\eqref{eq:unitarity} and Eq.~\eqref{eq:stability} and using the quartic self couplings given in Eqs.~\eqref{eqs:lambda}, we show the joint unitarity and stability constraints in Fig.~\ref{fig:UnitarityStability} for Type-I and Type-II 2HDM.
The best-fit points of $(\alpha\,,\beta)$ for Type-I and Type-II cases are used as in Table.~\ref{tab:BM_hH}.
As mentioned in the previous subsection, a fixed charged Higgs boson mass of $M_\pm=600\,\GeV$ is always taken in the Type-II case to evade the B-physics constraints.
For the best-fit points of $(\alpha,\beta)$ in Table.~\ref{tab:BM_hH}, a larger $m_{12}$ leads to a larger negative scalar quartic couplings $\lambda_1$ as indicated by Eq.~\eqref{eq:lam1}, therefore results in the perturbative unitarity (mostly from the $|a_-|\le$8$\pi$) and stability bounds on the  $m_{12}$, as depicted by the two panels of Fig.~\ref{fig:UnitarityStability}. 
For the fixed $m_{12}$, one can expect a larger positive $\lambda_3$ for larger heavy Higgs boson masses.
This breaks the perturbative unitarity through quartic coupling combination of $e_2$, therefore sets the upper bounds on the heavy Higgs boson masses for both Type-I and Type-II 2HDM. 
For the fixed heavy Higgs boson masses of $M_A=M_\pm$ in the Type-I case, a smaller $m_{12}$ leads to a relatively larger positive $\lambda_1$.
This results in a larger $|a_+|$, which in turn gives the lower bounds on $m_{12}$ on the left panel of Fig.~\ref{fig:UnitarityStability}.
The upper bounds to the mass mixing of $m_{12}$ in the Higgs potential set by the unitarity constraints and the stability constraints, which mainly come from the fact that the mass of the second $CP$-even Higgs boson $M_H$ is fixed. 
Correspondingly, the quartic scalar couplings of $\lambda_{1\,,2}$ are determined by $m_{12}$ for the best-fit points.
Since the $m_{12}$ enters into the Higgs cubic self couplings, we take their ranges to be
\beqn\label{eq:m12}
\textrm{Type-I} &:&20\lesssim m_{12}\lesssim 50\,\GeV \,,~~\textrm{with } M_A=M_\pm\in (200\,,280)\,\GeV \,,\non
\textrm{Type-II} &:& 0\leq m_{12}\lesssim 60\,\GeV\,,\non
&&\textrm{with } M_A\in (200\,,250)\,\GeV \textrm{ and } M_\pm = 600\,\GeV\,.
\eeqn
Here, we also limit the heavy Higgs boson masses of $(M_A\,, M_\pm)$ in the ranges that are consistent with the current LHC run-II searches for the $CP$-odd Higgs boson $A$ in the mass-degenerate scenario, as indicated in Fig.~\ref{fig:AhZexclu_LHC} below.

\begin{figure}[htb]
\centering
\includegraphics[width=0.49\textwidth]{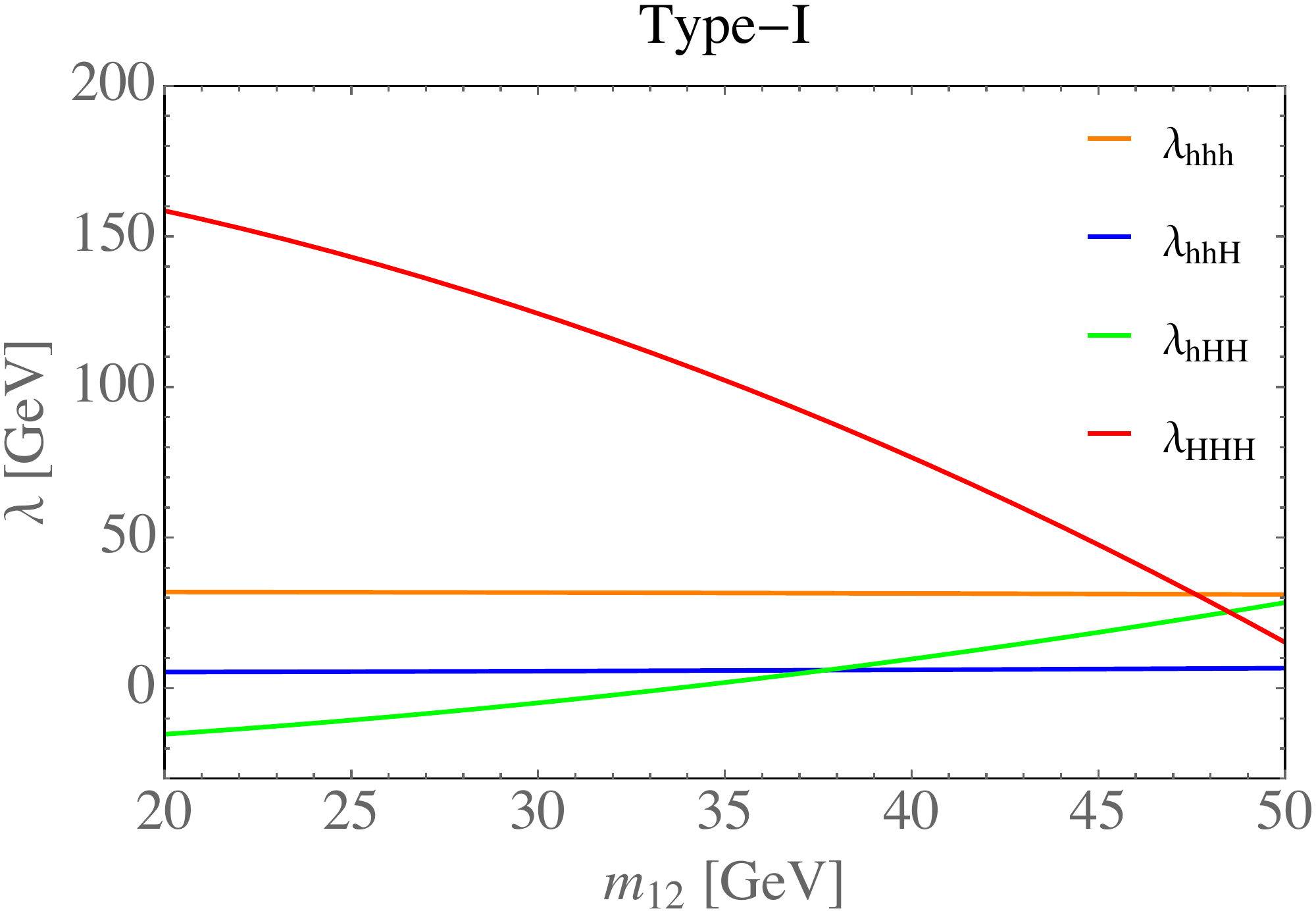}
\includegraphics[width=0.49\textwidth]{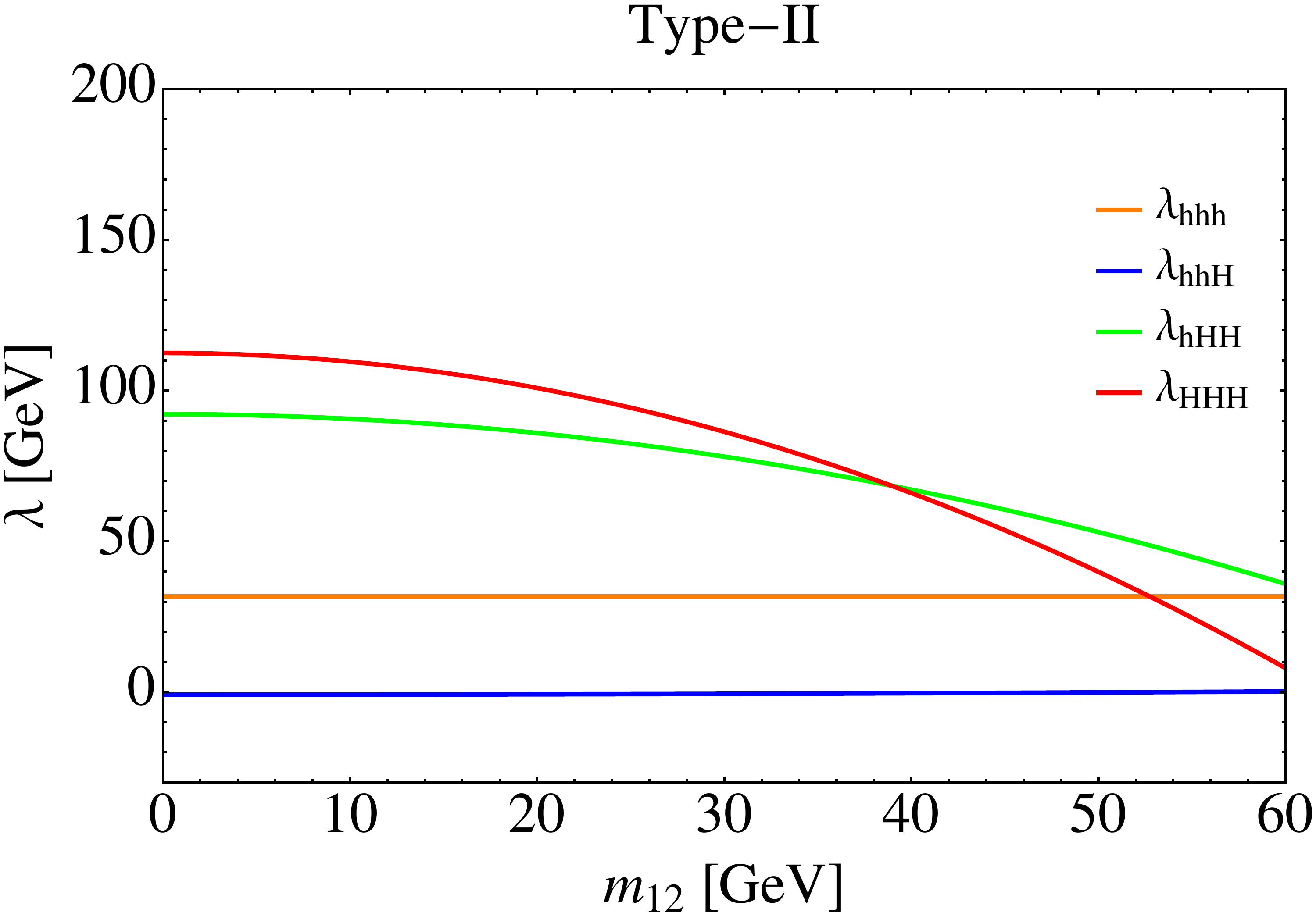}
\caption{\label{fig:lambda}
The dependences of the relevant Higgs cubic self couplings $\lambda_{hhh}$ (orange), $\lambda_{hhH}$ (blue), $\lambda_{hHH}$ (green), and $\lambda_{HHH}$ (red) on the soft mass term $m_{12}$ in the Type-I (left) and Type-II (right) 2HDM.
The best-fit points of $(\alpha\,,\beta)=(0.04\,,1.40)$ for Type-I case and $(\alpha\,,\beta)=(-0.25\,, 1.31)$ for Type-II case are taken as in Table.~\ref{tab:BM_hH}.
The ranges of $m_{12}$ are taken as in Eq.~\eqref{eq:m12} for Type-I and Type-II cases.
}
\end{figure}

The coefficients of the Higgs cubic self couplings in the physical basis will be used in the calculation of the direct Higgs pair productions at the LHC and the ILC, which are listed as follows
\beqs
\beqn
\lambda_{hhh}&=&  - \frac{1}{32 v s_\beta^2 c_\beta^2} \Big[ M_h^2 ( 3 s_{\alpha-\beta} + s_{3(\alpha-\beta)} - s_{3\alpha+\beta} - 3 s_{\alpha+3\beta} ) \non
&+& 4 m_{12}^2 ( c_{3\alpha - \beta} + c_{\alpha-3\beta} + 2 c_{\alpha+\beta})   \Big] \,,\\
\lambda_{hhH}&=& \frac{ c_{\alpha-\beta} }{2 v s_\beta c_\beta} \Big[ ( 2M_h^2 + M_H^2 ) s_\alpha c_\alpha + m_{12}^2 ( 1- 3 \frac{s_{2\alpha}}{ s_{2\beta} } ) \Big] \,,\\
\lambda_{hHH} &=& \frac{ s_{\beta-\alpha} }{2 v s_\beta c_\beta} \Big[ - ( M_h^2 + 2 M_H^2 ) s_\alpha c_\alpha + m_{12}^2 (1 + 3 \frac{ s_{2\alpha} }{ s_{2\beta} } )  \Big]  \,,\\
\lambda_{HHH}&=& - \frac{1}{32 v s_\beta^2 c_\beta^2 } \Big[ M_H^2 ( c_{3(\alpha-\beta)} - c_{3\alpha+ \beta} - 3 c_{\alpha-\beta} + 3 c_{\alpha+ 3\beta}  )  \non
&+& 4 m_{12}^2 ( s_{\alpha- 3\beta} - s_{3\alpha - \beta} + 2 s_{\alpha+ \beta}  )  \Big]\,.
\eeqn
\eeqs
In the alignment limit of $\beta-\alpha=\pi/2$, they become
\beqs
\beqn
\lambda_{hhh}&\to& \frac{M_h^2}{2v} \,,\\
\lambda_{hhH}&\to&0 \,,\\
\lambda_{hHH}&\to&  \frac{1}{2v} \Big( M_h^2 + 2 M_H^2 - 2 \frac{ m_{12}^2 }{ s_\beta c_\beta}  \Big) \,,\\
\lambda_{HHH}&\to& - \frac{1}{v\, t_{2\beta}} \Big(  M_H^2 - \frac{m_{12}^2}{ s_\beta c_\beta } \Big) \,.
\eeqn
\eeqs
In Fig.~\ref{fig:lambda}, we display the Higgs cubic self couplings versus the soft mass term $m_{12}$, with the unitarity/stability constraints in Eq.~\eqref{eq:m12} taken into account for two best-fit points listed in Table.~\ref{tab:BM_hH}. 
It turns out that the Higgs cubic self couplings of $\lambda_{hhh}$ are very close to the SM value of $\lambda_{hhh}^{\rm SM}\simeq 32\,\GeV$ in both Type-I and Type-II 2HDM.
The Higgs cubic self couplings of $\lambda_{hhH}$ are suppressed by the alignment parameter of $c_{\beta-\alpha}$.
With $c_{\beta-\alpha}=0.21$ in Type-I and $c_{\beta-\alpha}=0.01$ in Type-II 2HDM, $\lambda_{hhH}$ approaches to zero for the allowed range of $m_{12}$.
Jointly, one can envision that the cross sections of $\sigma[e^+ e^- \to hhZ]$ at the ILC $500\,\GeV$ run are almost independent of the $m_{12}$ inputs.
The Higgs cubic self couplings of $\lambda_{hHH}$ increase from $-15\,\GeV$ to $28\,\GeV$, with $m_{12}\in (20\,, 50)\,\GeV$ in the Type-I case; or decrease from $92\,\GeV$ to $36\,\GeV$, with $m_{12}\in (0\,,60)\,\GeV$ in the Type-II case.
Such behaviors are relevant to the relation between the $\sigma[e^+ e^- \to HHZ]$ at the ILC $500\,\GeV$ run and the $m_{12}$ inputs.
The Higgs cubic self couplings of $\lambda_{HHH}$ are always positive in both Type-I and Type-II cases.

\subsection{The constraints from the $CP$-odd Higgs boson $A$ searches}

Before we discuss the degenerate Higgs boson pair searches at the LHC, we impose the constraints via the LHC searches for the $CP$-odd Higgs boson $A$ in the mass-degenerate $h/H$ scenario. 
The decay modes and the corresponding partial decay widths of $CP$-odd Higgs boson $A$ are
\beqs\label{eqs:Awid}
\beqn
\Gamma[A\to gg]&=& \frac{G_F \alpha_s^2 \, M_A^3}{64\sqrt{2} \pi^3 } \Big| \sum_q \xi_A^q A_{1/2}^A(\tau_q)   \Big|^2\,,\\
\Gamma[A\to f \bar f]&=& \frac{G_F m_f^2 M_A}{4 \sqrt{2} \pi } N_{c\,,f}  (\xi_A^f)^2 \sqrt{ 1- \frac{4 m_f^2}{M_A^2} } \,,\label{eq:AffWid}\\
\Gamma[A\to hZ]&=&\frac{g^{2}c_{\beta-\alpha}^{2}}{64\pi M_{A}c_{W}^{2}}\lambda^{1/2} \Big( 1\,,\frac{m_{Z}^{2}}{M_{A}^{2}}\,,\frac{M_{h}^{2}}{M_{A}^{2}} \Big)\non
&\times&\Big[ m_{Z}^{2}-2(M_{A}^{2}+M_{h}^{2})+\frac{(M_{A}^{2}-M_{h}^{2})^{2}}{m_{Z}^{2}}  \Big]\,,\label{eq:AhZWid}\\
\Gamma[A\to HZ]&=&\frac{g^{2}s_{\beta-\alpha}^{2}}{64\pi M_{A}c_{W}^{2}}\lambda^{1/2} \Big( 1\,,\frac{m_{Z}^{2}}{M_{A}^{2}}\,,\frac{M_H^{2}}{M_{A}^{2}} \Big)\non
&\times&\Big[ m_{Z}^{2}-2(M_{A}^{2}+M_H^{2})+\frac{(M_{A}^{2}-M_H^{2})^{2}}{m_{Z}^{2}}  \Big]\,,\label{eq:AHZWid}
\eeqn
\eeqs
with $N_{c, f}=3\,(1)$ for quarks (leptons).
The three-body phase space factor reads
\beqn\label{eq:threebody_phase}
\lambda^{1/2}(1\,,x^2 \,, y^2)&\equiv& \Big[ ( 1-x^2-y^2 )^2 - 4x^2 y^2  \Big]^{1/2}\,.
\eeqn
For the mass-degenerate $h/H$ scenario, one cannot discriminate two decay channels of $A\to hZ$ and $A\to HZ$ for specific final states, such as $b \bar b + \ell^+ \ell^-$.
The signal rates for this scenario should be evaluated as
\beqn
\sigma_{\rm tot}&=& \sigma[pp\to AX] \times \Big( \textrm{BR}[A\to hZ] \times \textrm{BR}[h\to b \bar b]\non
&+& \textrm{BR}[A\to HZ] \times  \textrm{BR}[H\to b \bar b]   \Big)\times\textrm{BR}[Z\to \ell^+ \ell^-]\,.
\eeqn

\begin{figure}[htb]
\centering
\includegraphics[width=0.6\textwidth]{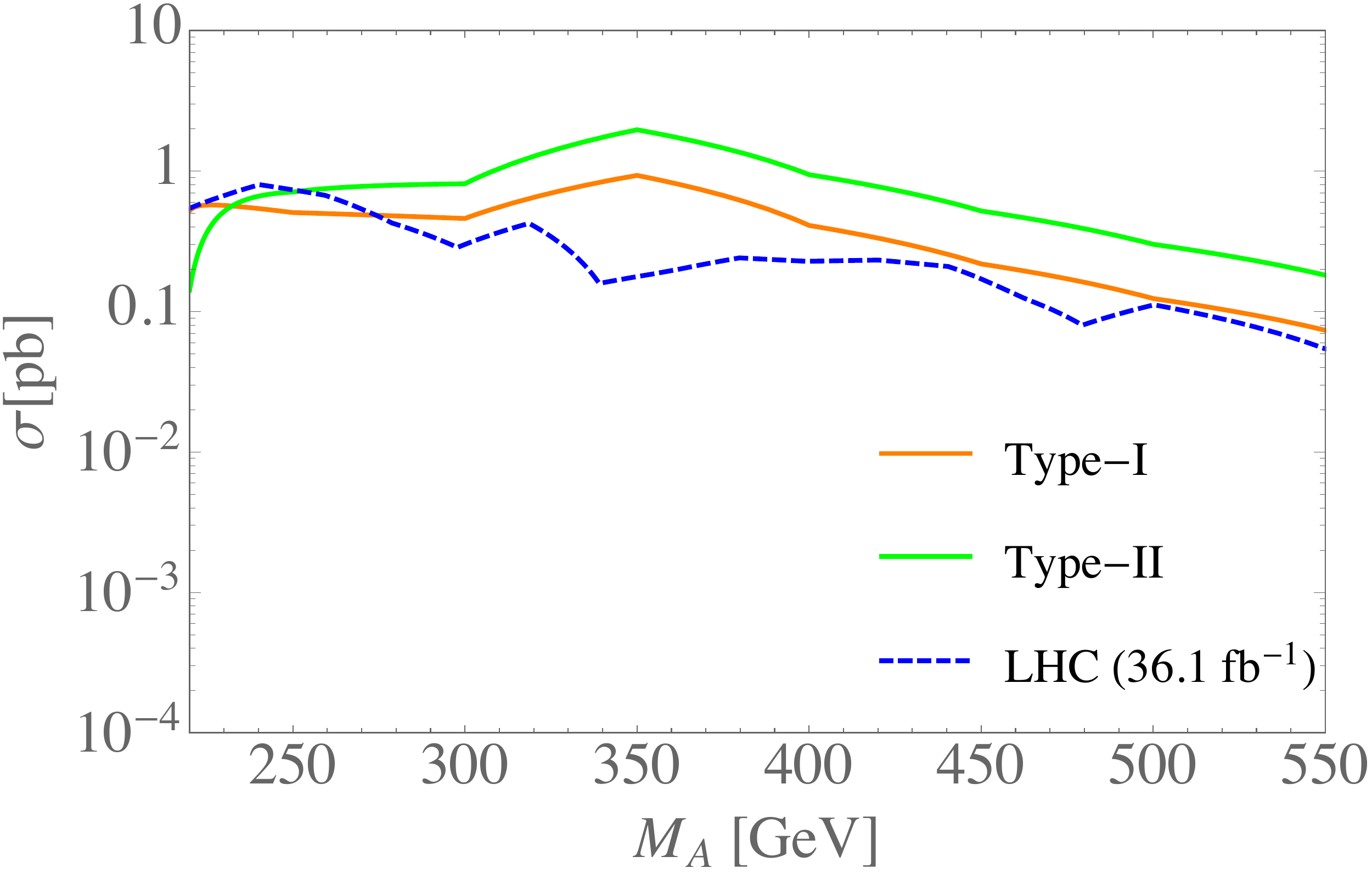}
\caption{\label{fig:AhZexclu_LHC}
The current LHC exclusion limits~\cite{Aaboud:2017cxo} to the $h/H$ degenerate case through the $A\to hZ\to b\bar b + \ell^+ \ell^-$ final states, with an integrated luminosity of $36.1\,\fb^{-1}$ (blue dashed line).
The best-fit points of $(\alpha\,,\beta)=(0.04\,,1.40)$ for Type-I case and $(\alpha\,,\beta)=(-0.25\,, 1.31)$ for Type-II case are taken as in Table.~\ref{tab:BM_hH}.
}
\end{figure}

The LHC searches for a $CP$-odd Higgs boson via the $A\to hZ\to b \bar b + \ell^+ \ell^-$ were previous carried out by both ATLAS~\cite{Aad:2015wra} and CMS~\cite{Khachatryan:2015lba} collaborations at the $8\,\TeV$ run.
The most recent results from the ATLAS searches at the LHC $13\,\TeV$ run with integrated luminosity of $36.1\,\fb^{-1}$ is given in Ref.~\cite{Aaboud:2017cxo}. 
We estimate the production cross sections of $\sigma(gg\to A)$ at the NLO by using the package of {\tt SusHi}~\cite{Harlander:2012pb}, by using the best-fit points as in Table.~\ref{tab:BM_hH}. 
From Fig.~\ref{fig:AhZexclu_LHC}, we find that the $h/H$ mass-degenerate scenario for either Type-I or Type-II has been tightly constrained by the current exclusion limits from the LHC $13\,\TeV$ run with an integrated luminosity of $36.1\,\fb^{-1}$.
Combining with the joint perturbative unitarity and stability bounds shown in Fig.~\ref{fig:UnitarityStability}, the $h/H$ mass-degenerate scenario can still exist in the mass ranges of $M_A\lesssim 280\,\GeV$ for Type-I 2HDM and $M_A\lesssim 260\,\GeV$ for Type-II 2HDM, respectively.
Previously, we further restrict the mass ranges of $M_A$ in Eq.~\eqref{eq:m12}, in order to maximize our parameter choices of $m_{12}$ for the Higgs pair productions.
It does not mean the mass ranges of $M_A$ in Eq.~\eqref{eq:m12} are constrained by the current LHC search result.
The specific mass of the $CP$-odd Higgs boson $A$ will not play any role in our discussion of the future experimental tests below.
On the other hand, the future searches for the $CP$-odd Higgs boson $A$ via this channel at the LHC $14\,\TeV$ run will play a decisive role in justifying or falsifying this scenario.


\section{The LHC searches for degenerate Higgs bosons: constraints and pair productions}
\label{section:Hpair_LHC}

\subsection{The total cross section of gluon-gluon fusion to Higgs pairs}

By using the best-fit points in Table.~\ref{tab:BM_hH} and the range of $m_{12}$ in Eq.~\eqref{eq:m12} after the set of constraints, we are ready to present the main results of the Higgs pair productions at the LHC.
The cross sections we need to evaluate are
\beqn
&&\sigma[gg\to hh]\,,\qquad  \sigma[gg\to hH]\,,\qquad \sigma[gg\to HH]\,,
\eeqn
where the individual cross section of $\sigma[gg\to h_i h_j]$ was first obtained in Ref.~\cite{Plehn:1996wb} for both SM and MSSM cases.
The differential cross section at the LO reads
\beqn\label{eq:dsigma_ggtohh_parton}
\frac{d \hat \sigma}{d \hat t}&=&c^{ij} \frac{G_F^2 \alpha_s^2}{256\, (2\pi)^3}\Big\{  \Big| \sum_{q=t\,,b} (C_\triangle^{ij} F_\triangle + C_\Box^{ij} F_\Box )  \Big|^2 + \Big| \sum_{q=t\,,b} C_\Box^{ij} G_\Box \Big|^2 \Big\}\,,
\eeqn
with $c^{ij}=1/2\,(1)$ for $i=j\,(i\neq j)$.
$C_\triangle$ and $C_\Box$ represent the coefficients of the triangle and box diagrams, respectively.
The Higgs cubic self couplings contribute to the $C_\triangle$'s, and they read
\beqs
\beqn
C_\triangle^{hh}&=& \frac{ 3\lambda_{hhh}v\, \xi_h^q }{\hat s - M_h^2 + i M_h \Gamma_h} + \frac{2 \lambda_{hhH} v\, \xi_H^q }{ \hat s - M_H^2 + i M_H \Gamma_H}\,,\\
C_\triangle^{hH}&=&\frac{ 2\lambda_{hhH}v\, \xi_h^q }{\hat s - M_h^2 + i M_h \Gamma_h} + \frac{2 \lambda_{hHH} v\, \xi_H^q }{ \hat s - M_H^2 + i M_H \Gamma_H}\,,\\
C_\triangle^{HH}&=& \frac{ 2\lambda_{hHH}v\, \xi_h^q }{\hat s - M_h^2 + i M_h \Gamma_h} + \frac{3 \lambda_{HHH} v\, \xi_H^q}{ \hat s - M_H^2 + i M_H \Gamma_H}\,.
\eeqn
\eeqs
The coefficients of $C_\Box$ are determined by the dimensionless Yukawa couplings of the Higgs bosons
\beqn
&& C_\Box^{hh}= (\xi_h^q)^2\,,\qquad C_\Box^{hH}=\xi_h^q \xi_H^q\,,\qquad C_\Box^{HH}=(\xi_H^q)^2\,,
\eeqn
for $q=(t\,,b)$.
The form factors of the triangle and box diagrams $(F_\triangle\,, F_\Box\,, G_\Box)$ in Eq.~\eqref{eq:dsigma_ggtohh_parton} can be found in the Appendix of Ref.~\cite{Plehn:1996wb}.
The asymptotic behaviors of these form factors in the large quark mass and small quark mass limits read
\beqs\label{eqs:formfactor}
\beqn
m_q^2\gg \hat s&:& F_\triangle\simeq \frac{2}{3}+ \mO(\frac{\hat s }{m_q^2})\,,\qquad F_\Box\simeq - \frac{2}{3}+ \mO(\frac{\hat s }{m_q^2})\,,\qquad G_\Box \simeq \mO(\frac{\hat s }{m_q^2})\,,\\
m_q^2\ll \hat s&:& F_\triangle\simeq - \frac{m_q^2}{\hat s} \Big[ \log\Big( \frac{m_q^2}{\hat s} \Big)  + i\pi \Big]^2+ \mO(\frac{m_q^2}{\hat s } )\,,\qquad F_\Box\,, G_\Box \simeq \mO(\frac{ m_q^2 }{\hat s})\,.
\eeqn
\eeqs
In practice, we evaluate the corresponding Passarino-Veltman (PV) integrals are evaluated by using the {\tt LoopTools} package~\cite{Hahn:1998yk}.

\begin{figure}[htb]
\centering
\includegraphics[width=0.7\textwidth]{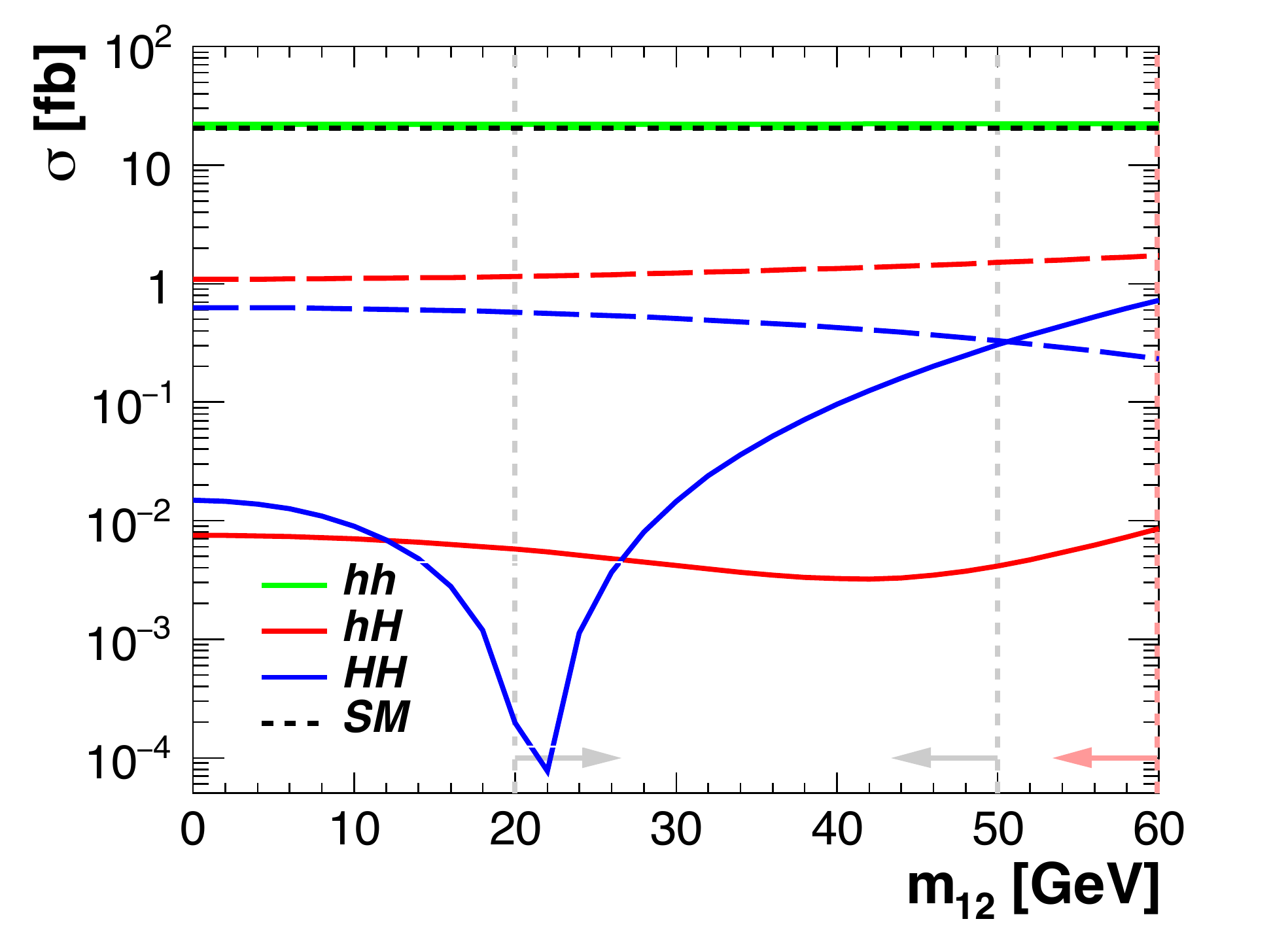}
\caption{\label{fig:Hpair_total}
The LO cross sections of $\sigma[gg\to hh]$ (green), $\sigma[gg\to hH]$ (red), and $\sigma[gg\to HH]$ (blue) versus the soft mass $m_{12}$ in both Type-I (solid) and Type-II (dashed) 2HDM at the LHC $\sqrt{s}=14\,\TeV$ run.
The best-fit points of $(\alpha\,,\beta)=(0.04\,,1.40)$ for Type-I case and $(\alpha\,,\beta)=(-0.25\,, 1.31)$ for Type-II case are taken as in Table.~\ref{tab:BM_hH}.
In comparison, the LO cross section of $\sigma[gg\to h_{\rm SM} h_{\rm SM}]=20.5\,\fb$ (black dotted) is listed as well. Note that the allowed regions obtained from theoretical constraints in $m_{12}$ are also indicated by the gray and pink arrows for Type-I and Type-II respectively.
}
\end{figure}

At the LHC, the differential cross section in the lab frame is obtained by convoluting the parton-level cross section in Eq.~\eqref{eq:dsigma_ggtohh_parton} with the gluon PDFs
\beqn\label{eq:dsigma_ggtohh}
\frac{d^2 \sigma}{d M_{hh} dp_T}&=& \int_\tau^1 \, \frac{dx}{x} f_{g}(x\,, \mu_F) f_g(\frac{\tau}{x}\,,\mu_F) \frac{2 M_{hh} }{s} \frac{d\hat \sigma}{d p_T}\,,
\eeqn
where $s$ is the squared center-of-mass energy at the LHC, $M_{hh}$ is the invariant mass of the Higgs pairs, $\tau= M_{hh}^2/s$, and $p_T$ denotes the transverse momentum of the Higgs boson.
In practice, we use the LO MSTW PDF~\cite{Martin:2009iq} for the evaluation.
The LO cross sections of individual production mode of $\sigma[gg\to hh]$, $\sigma[gg\to hH]$, and $\sigma[gg\to HH]$ are displayed in Fig.~\ref{fig:Hpair_total}, with the renormalization and factorization scale set to be $\mu_R=\mu_F=M_{hh}$.
The cross sections of $\sigma[gg\to hh]$ are very close to the corresponding SM predicted value, as stated previously.
The contributions from $\sigma[gg\to hH]$ and $\sigma[gg\to HH]$ are sub-leading ones, yet they play a role in determining the signal rates.
A dip is shown in the cross section of $\sigma[gg\to hH ]$ versus $m_{12}$, which roughly matches the position where the Higgs cubic self coupling $\lambda_{hHH}$ flips sign.
The next-to-leading order contributions to the Higgs pair productions at the LHC are known to be significant~\cite{Dawson:1998py, Shao:2013bz, deFlorian:2013uza, deFlorian:2013jea, Borowka:2016ehy}.
Our estimation below focus on the future experimental significances via various channels, where a same $K$-factor can be assumed for both $h/H$ mass-degenerate case and the SM Higgs boson case, as in Ref.~\cite{Barger:2014taa}.
Therefore, the LO results are sufficient for our estimation below.

\subsection{The cross sections of Higgs pairs to various final states}

Next, we proceed to present the cross sections of the Higgs pairs to specific final states with the mass-degenerate Higgs benchmark points in Table~\ref{tab:BM_hH}.
Two leading final states of $4b$ and $2b\,2\gamma$ will be taken into account~\cite{Cadamuro_talk}. 
The signal rates are estimated as follows
\beqn\label{eq:hhsignal}
\sigma[gg\to(hh\,,hH\,,HH)\to (XXYY)]&=& \sigma[gg\to hh] (\kappa_{XY}{\rm Br}[h\to XX] {\rm Br}[h\to YY])\non
&+&\sigma[gg\to hH]\Big( {\rm Br}[h\to XX] {\rm Br}[H\to YY]+(h\leftrightarrow H) \Big) \non
&+&\sigma[gg\to HH] (\kappa_{XY}{\rm Br}[H\to XX] {\rm Br}[H\to YY])\,,\non
\eeqn
with $\kappa_{XY}=1\,(2)$ for $X= Y\,(X\neq Y)$.
For a single SM-like Higgs boson with mass $\sim 125\,\GeV$, the ratio of signal rates between the $4b$ and $2b\, 2\gamma$ final states is fixed to be $\sigma[4b] :\sigma[2b\, 2\gamma]\approx 127:1$.
This always holds, no matter how one modifies the SM-like Higgs cubic self couplings and includes the additional resonance contributions.
We find the ratios of signal rates between these two channels are generally different from the single SM-like Higgs boson case, which reads $\sigma[4b]: \sigma[2b\, 2\gamma]\approx 140:1 $ with $m_{12}=50\,\GeV$ in the Type-I case, and $\sigma[4b]: \sigma[2b\, 2\gamma]\approx 129:1$ with $m_{12}=60\,\GeV$ in the Type-II case.

\begin{figure}[htb]
\centering
\includegraphics[width=0.49\textwidth]{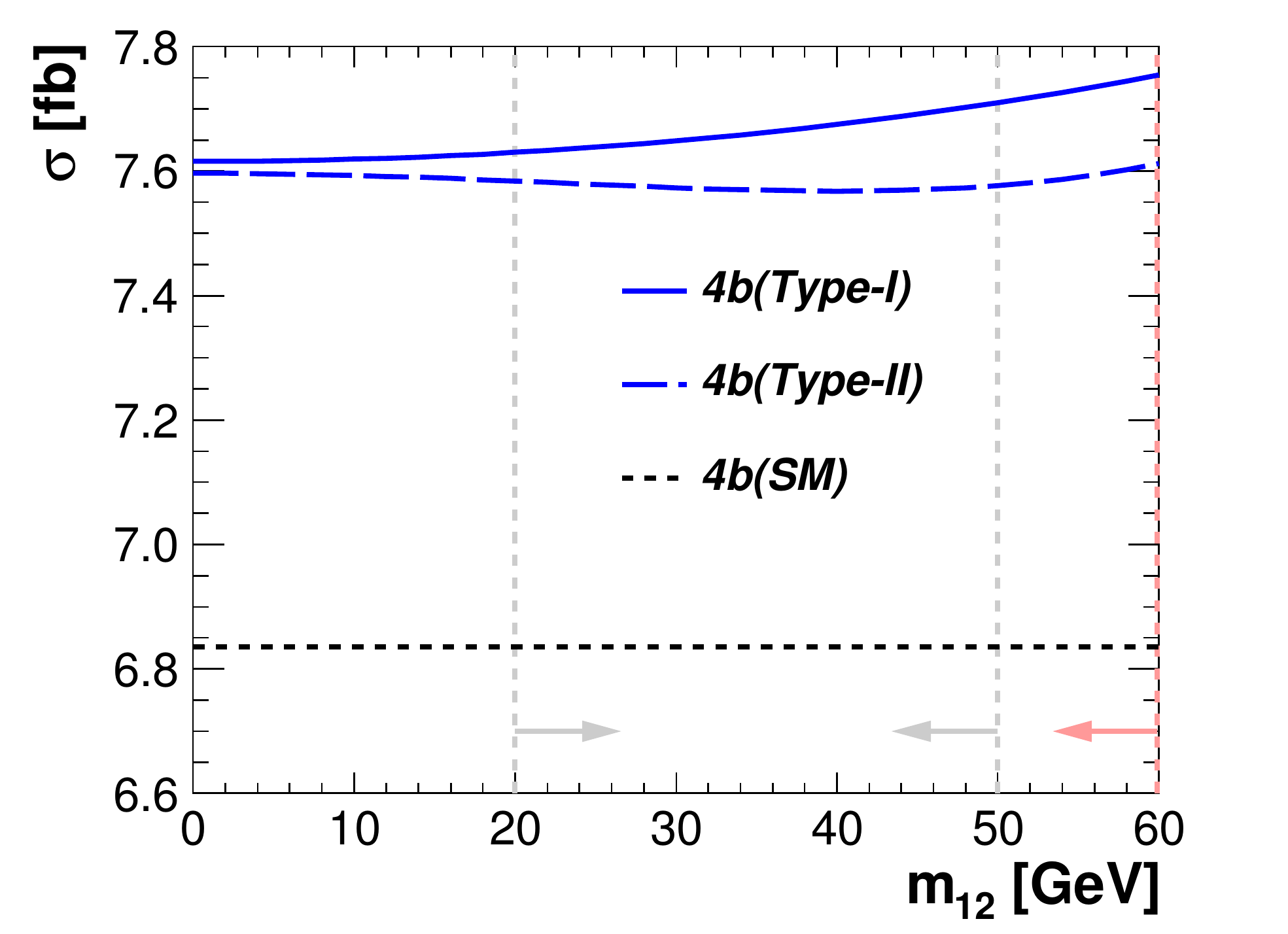}
\includegraphics[width=0.49\textwidth]{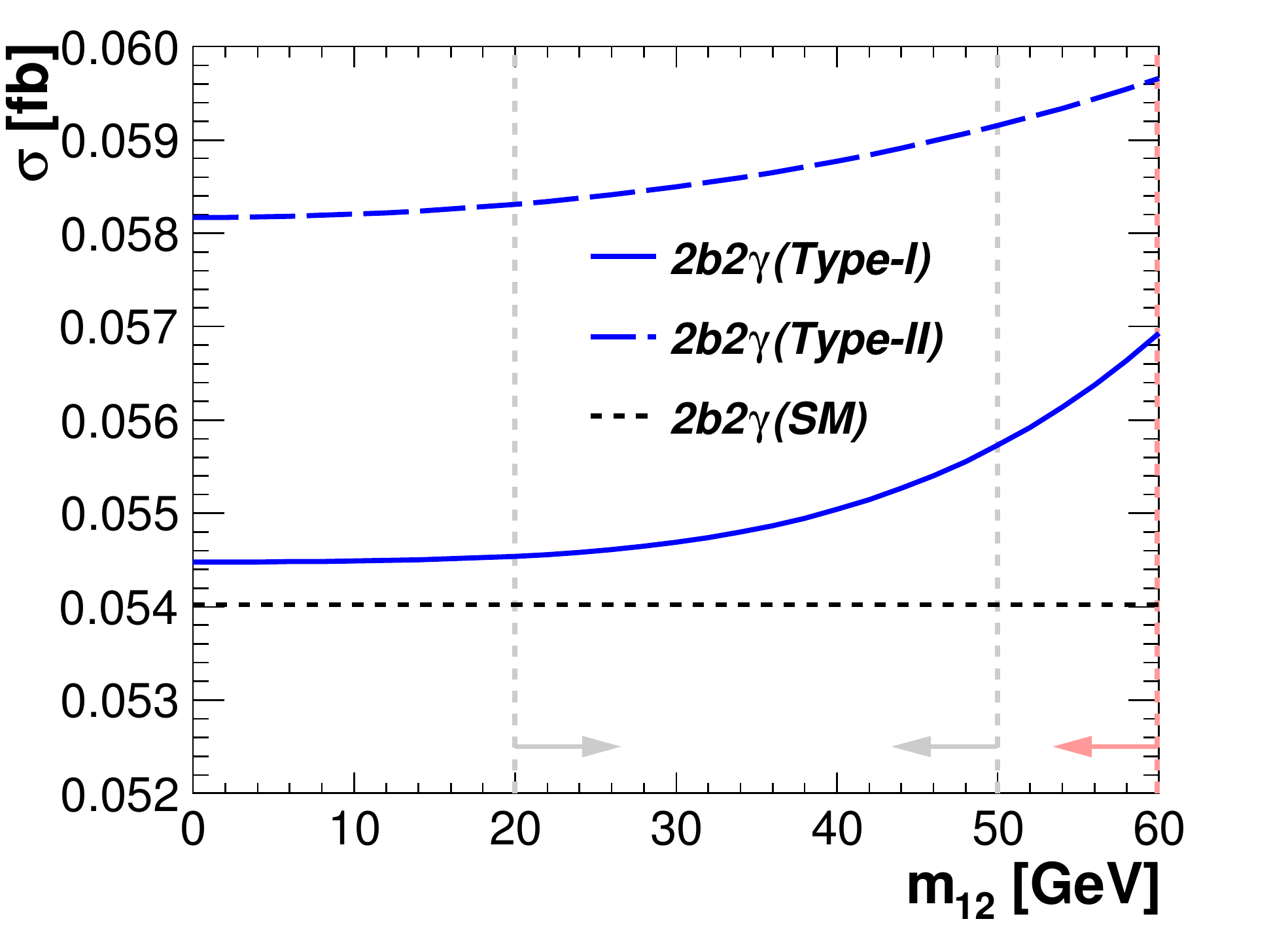}
\caption{\label{fig:Hpair_channel}
The LHC $14\,\TeV$ LO cross sections of $\sigma[gg\to (hh\,,hH\,,HH)]$ to final states of $4b$ (left panel) and $2b\, 2\gamma$ (right panel) versus the soft mass $m_{12}$ in the Type-I (solid lines) and Type-II (dashed lines) 2HDM.
The best-fit points of $(\alpha\,,\beta)=(0.04\,,1.40)$ for Type-I case and $(\alpha\,,\beta)=(-0.25\,, 1.31)$ for Type-II case are taken as in Table.~\ref{tab:BM_hH}.
In comparison, the LO cross sections of $\sigma[gg\to h_{\rm SM} h_{\rm SM}]$ to the corresponding final states (dotted lines) are displayed as well. Note that the allowed regions obtained from theoretical constraints in $m_{12}$ are also indicated by the gray and pink arrows for Type-I and Type-II respectively.
}
\end{figure}

The LHC $14\,\TeV$ cross sections for $\sigma[gg\to (hh\,,hH\,,HH)\to (4b\,, 2b2\gamma)]$ final states versus the soft mass term $m_{12}$ are shown in Fig.~\ref{fig:Hpair_channel}.
We find enhancements of both $4b$ and $2b\,2\gamma$ signal rates, in comparison to the SM Higgs boson pair productions. 
This kind of enhancements were previously investigated in Ref.~\cite{Han:2015pwa}. 
However, the parameter region of $m_{12}$ is severely restricted by the theoretical constraints as shown in Fig.~\ref{fig:UnitarityStability}.
The LO cross sections for the $4b$ final states are moderately enhanced to $\sim7.7\,\fb$ with $m_{12}=50\,\GeV$ in the Type-I case, or $\sim 7.6\,\fb$ with $m_{12}=60\,\GeV$ in the Type-II case.
The corresponding LO cross section for the $4b$ channel in the SM is $\simeq 6.8\,\fb$.
By extrapolating the current LHC run-II results from $13\,\TeV$, the ATLAS and CMS give conservative estimations to the significances for the SM Higgs boson pairs at the HL LHC runs as~\cite{Cadamuro_talk}
\beqn
\begin{array}{rclc}
{\rm ATLAS}: & 1.05\,\sigma & \textrm{for} & 2b\, 2\gamma \\
{\rm CMS}: & (0.39\,\sigma\,,1.6\,\sigma) & \textrm{for} & (4b\,,2b\,2\gamma)
\end{array}
\eeqn
We summarize the significance of the Type-I and Type-II $h/H$ mass-degenerate Higgs boson pairs via the $4b$ and $2b\, 2\gamma$ channels in Table.~\ref{tab:hh_LHC}.
We note that the current LHC run-II results are not optimal for the HL-LHC runs, future improvements to the significance via the $2b\, 2\gamma$ channel should be expected.

\begin{table}[htb]
\begin{center}
\begin{tabular}{c|c|c}
\hline\hline
  & $4 b$ & $2b\, 2\gamma$   \\\hline
  Type-I $(m_{12})$  &  $\sim 0.44\,\sigma\, (20-50\,\GeV)$  & $1.61\,\sigma\, (20\,\GeV)~~1.64\,\sigma\, (50\,\GeV)$  \\ \hline
  Type-II $(m_{12})$  &  $\sim 0.43\,\sigma\, (0-60\,\GeV)$  & $1.71\,\sigma\, (0\,\GeV)~~1.75\,\sigma\, (60\,\GeV)$  \\ \hline\hline 
\end{tabular}
\caption{
The significances of the $h/H$ mass-degenerate Higgs boson pair production measurements via the $4b$ and $2b\, 2\gamma$ channels at the HL-LHC.
}
\label{tab:hh_LHC}
\end{center}
\end{table}


\section{Probes of degenerate Higgs bosons at the $e^+ e^-$ colliders}
\label{section:Hpair_ee}

The future plans of the high-energy $e^+ e^-$ colliders include the CEPC~\cite{CEPC-SPPCStudyGroup:2015csa}, ILC~\cite{Fujii:2015jha}, and TLEP~\cite{Gomez-Ceballos:2013zzn}.
They will play a role as Higgs factory to produce millions of SM-like Higgs bosons for the precise measurements, with the running at center-of-mass energy of $\sqrt{s}=250\,\GeV$, which will provide excellent opportunity for us to examine the Higgs properties in many NP models~\cite{Gu:2017ckc}.
It was pointed out in Ref.~\cite{McCullough:2013rea} that the precise measurement of the Higgs production cross section at the CEPC or ILC is sensitive to the Higgs self couplings at the NLO.
At the ILC, it is likely to upgrade the center-of-mass energy up to $\sqrt{s}=500\,\GeV$, so that it can directly produce $125\,\GeV$ Higgs boson pairs associated with a $Z$-boson.

\subsection{The CEPC measurements of the degenerate Higgs scenario}

The circular electron-positron collider (CEPC) will operate at the center-of-mass energy of $\sqrt{s}=250\,\GeV$.
A key physical goal of CEPC is to measure the Higgs boson mass precisely, which can be as small as $\sim 5.9\,\MeV$ with an integrated luminosity of $5\,\ab^{-1}$. 
The precision on $\sigma(ZH)$ is about 0.51\% combining all decay modes for the $Z$-boson with the same luminosity~\cite{CEPC-SPPCStudyGroup:2015csa}.
The resolution for the recoil mass peak measurement is about 400 MeV for ILC~\cite{Behnke:2013xla,ILC-TDR,Behnke:2013lya}.

With the integrated luminosity of $5\,\ab^{-1}$, CEPC can measure the cross section of $\sigma[e^+ e^- \to hZ]$ with an accuracy of $\sim 0.51\,\%$~\cite{CEPC-SPPCStudyGroup:2015csa}, by combining both the leptonic and hadronic decay modes of $Z$-bosons.
The LO cross section of the SM Higgs boson production associated with $Z$-boson at the CEPC reads $\sigma[e^+ e^- \to h_{\rm SM} Z]\simeq 221.54\,\fb$~\footnote{See Refs.~\cite{Sun:2016bel,Gong:2016jys,Fleischer:1982af,Kniehl:1991hk,Denner:1992bc} for more precisely prediction including higher order corrections.}.
The cross sections for the best-fit points in Table.~\ref{tab:BM_hH} are $\sigma[e^+ e^- \to h Z]\simeq 211.55\,\fb$ and $\sigma[e^+ e^- \to HZ] \simeq 9.99\,\fb$ for the Type-I case, $\sigma[e^+ e^- \to h Z]\simeq 221.52\,\fb$ and $\sigma[e^+ e^- \to HZ] \simeq 0.02\,\fb$ for the Type-II case.
Combined with the leading decay modes in Table.~\ref{tab:BM_hH}, we have $\sigma[e^+ e^- \to h(\to b \bar b) Z]\simeq 124.65\,\fb$ and $\sigma[e^+ e^- \to H (\to b \bar b) Z] \simeq 0.26\,\fb$ for the Type-I case, or $\sigma[e^+ e^- \to h(\to b \bar b) Z]\simeq 124.87 \,\fb$ and $\sigma[e^+ e^- \to H(\to b \bar b) Z]\simeq 1.8\times 10^{-2}\,\fb$ for the Type-II case.
Due to the jet energy resolutions, one does not expect to distinguish two separate peaks from two mass-degenerate Higgs bosons, with mass split of $\sim 0.1\,\GeV$.
Instead, the inclusive cross sections of $h/H$ mass-degenerate Higgs bosons with $b \bar b$ final states are about $\sim 2\,\%$ lower than the SM predicted values for both Type-I and Type-II cases. Compared with $\sim 0.28\,\%$ precision that could be achieved at CEPC~\cite{CEPC-SPPCStudyGroup:2015csa}, there will be roughly $7\,\sigma$ deviation.  
Therefore, a decrease of the cross section for the $\sigma[e^+ e^- \to hZ]$ will be a first indication from the best-fit points in Table.~\ref{tab:BM_hH}.

\subsection{The direct probes of the degenerate Higgs scenario at the ILC}

\begin{figure}[!thb]
\centering
\includegraphics[width=0.49\textwidth,trim=70 560 30 35, clip]{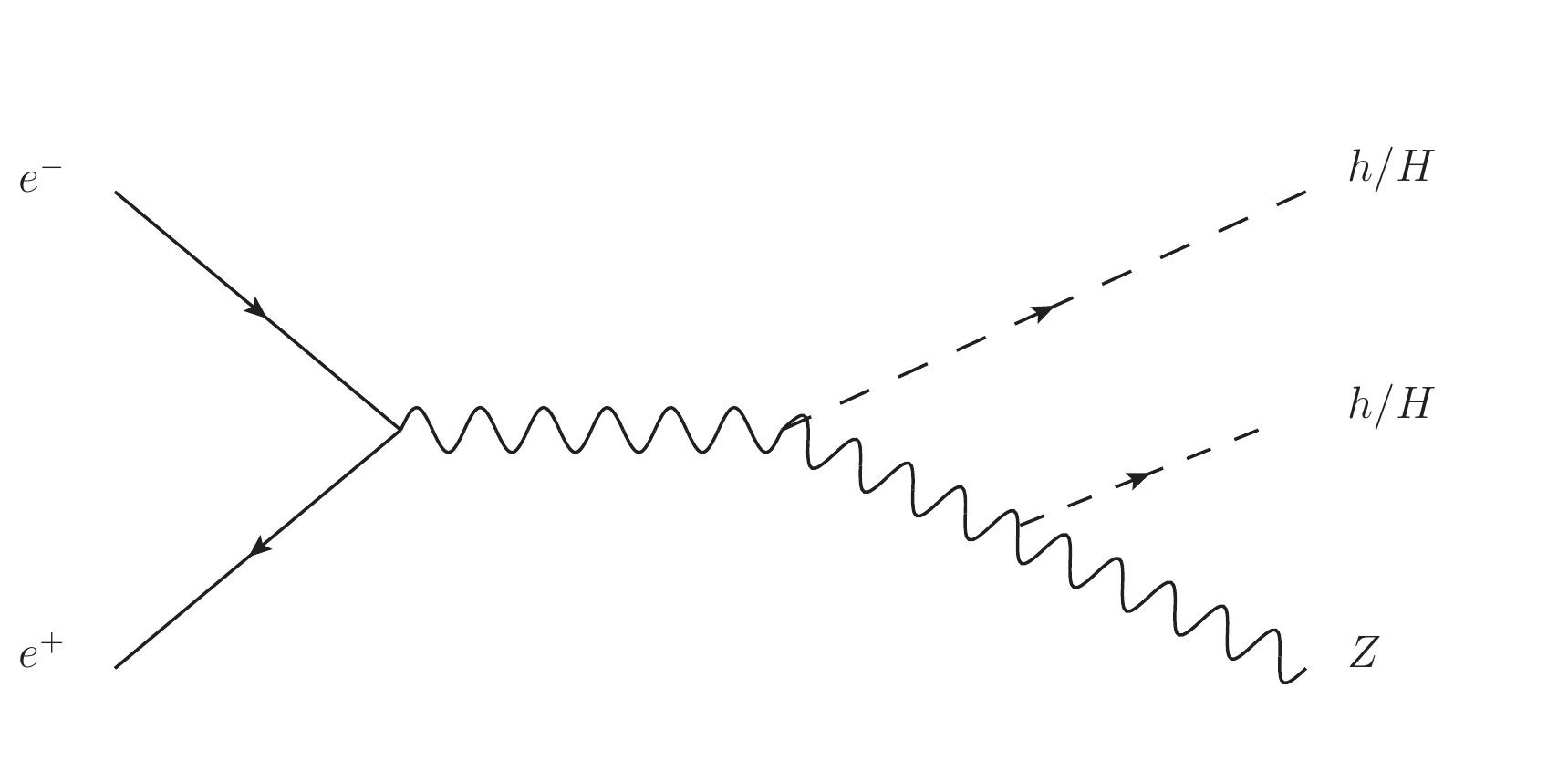}
\includegraphics[width=0.49\textwidth,trim=70 560 30 35, clip]{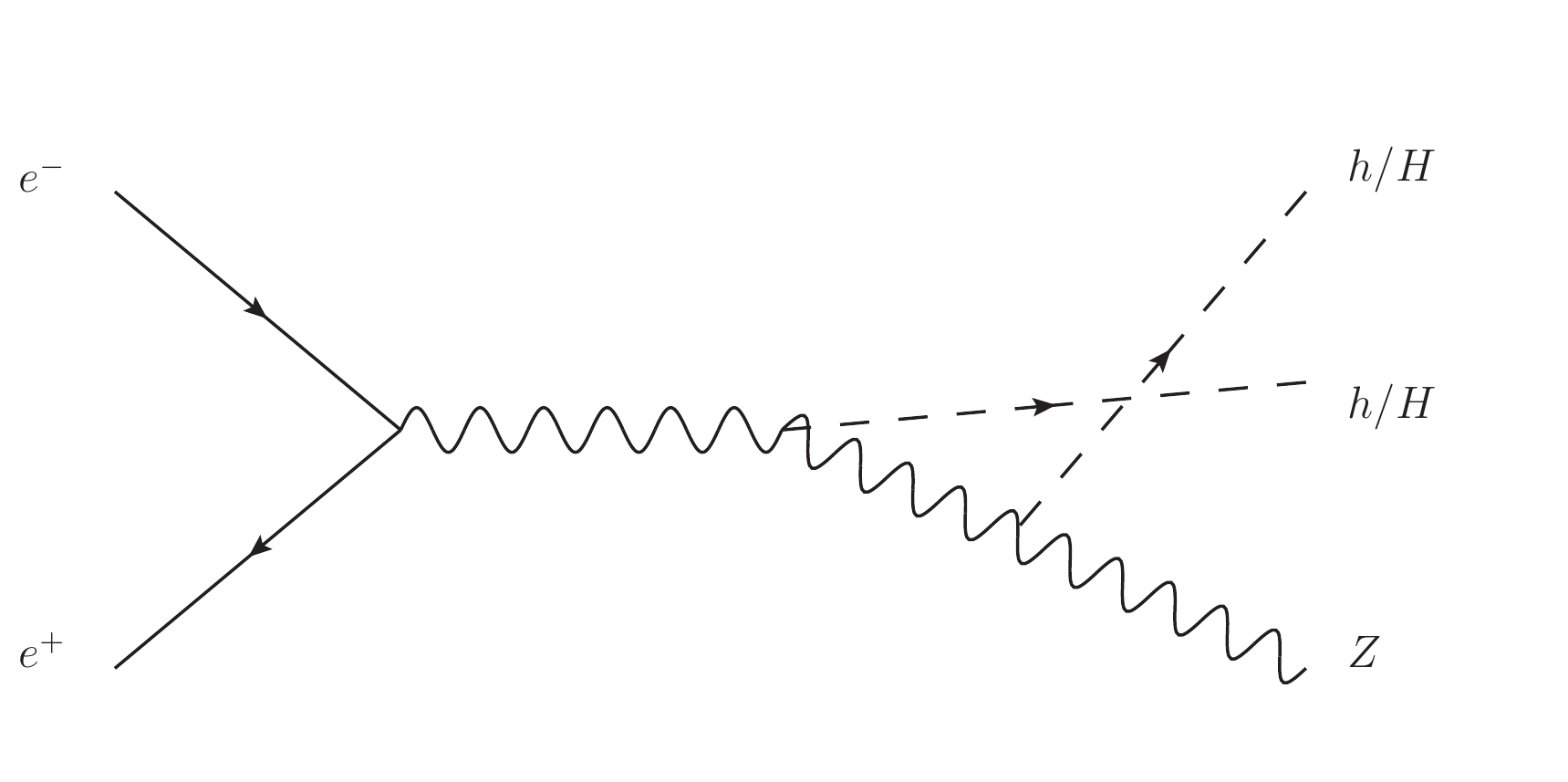}
\includegraphics[width=0.49\textwidth,trim=70 560 30 35, clip]{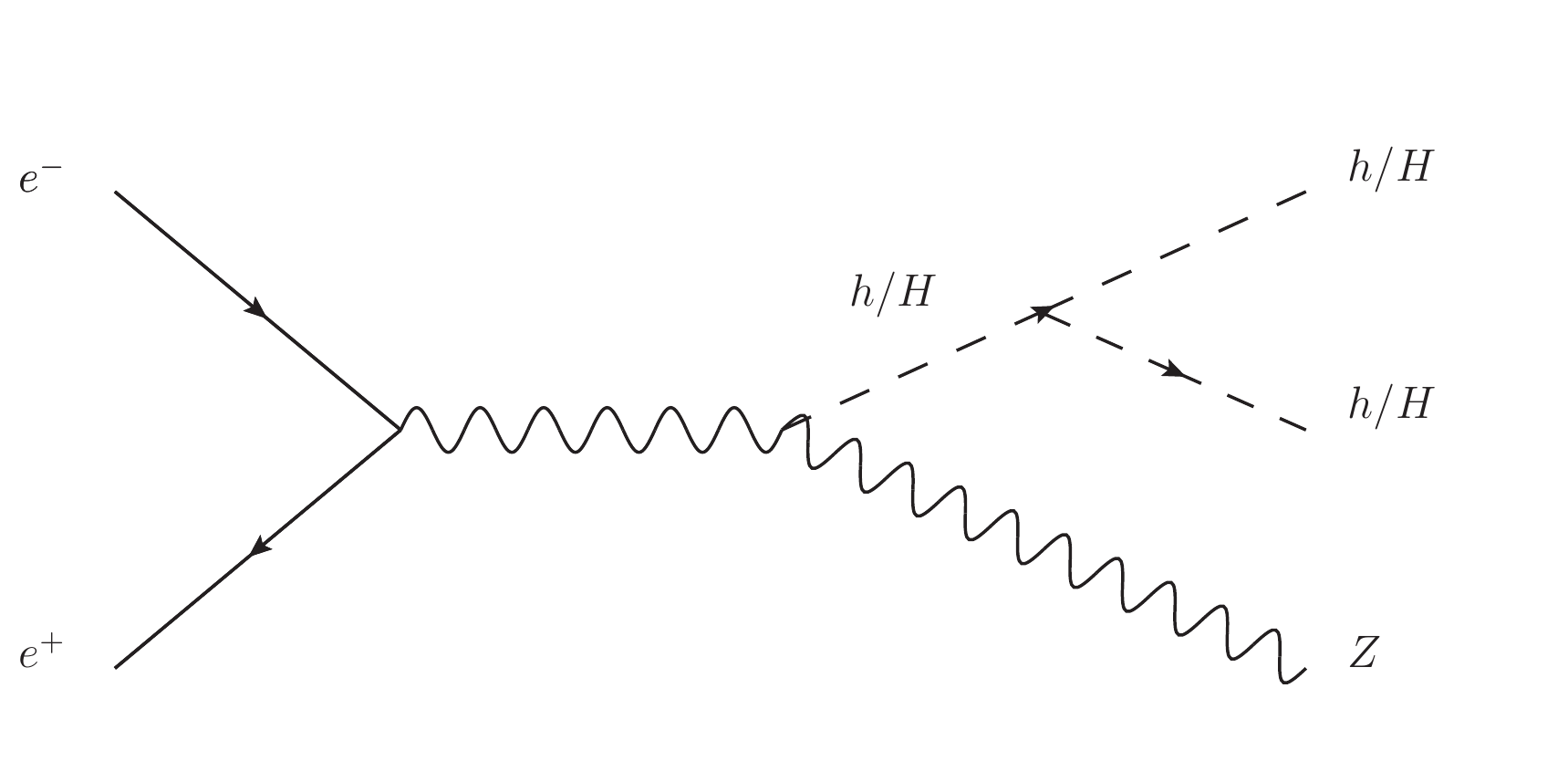}
\includegraphics[width=0.49\textwidth,trim=70 560 30 35, clip]{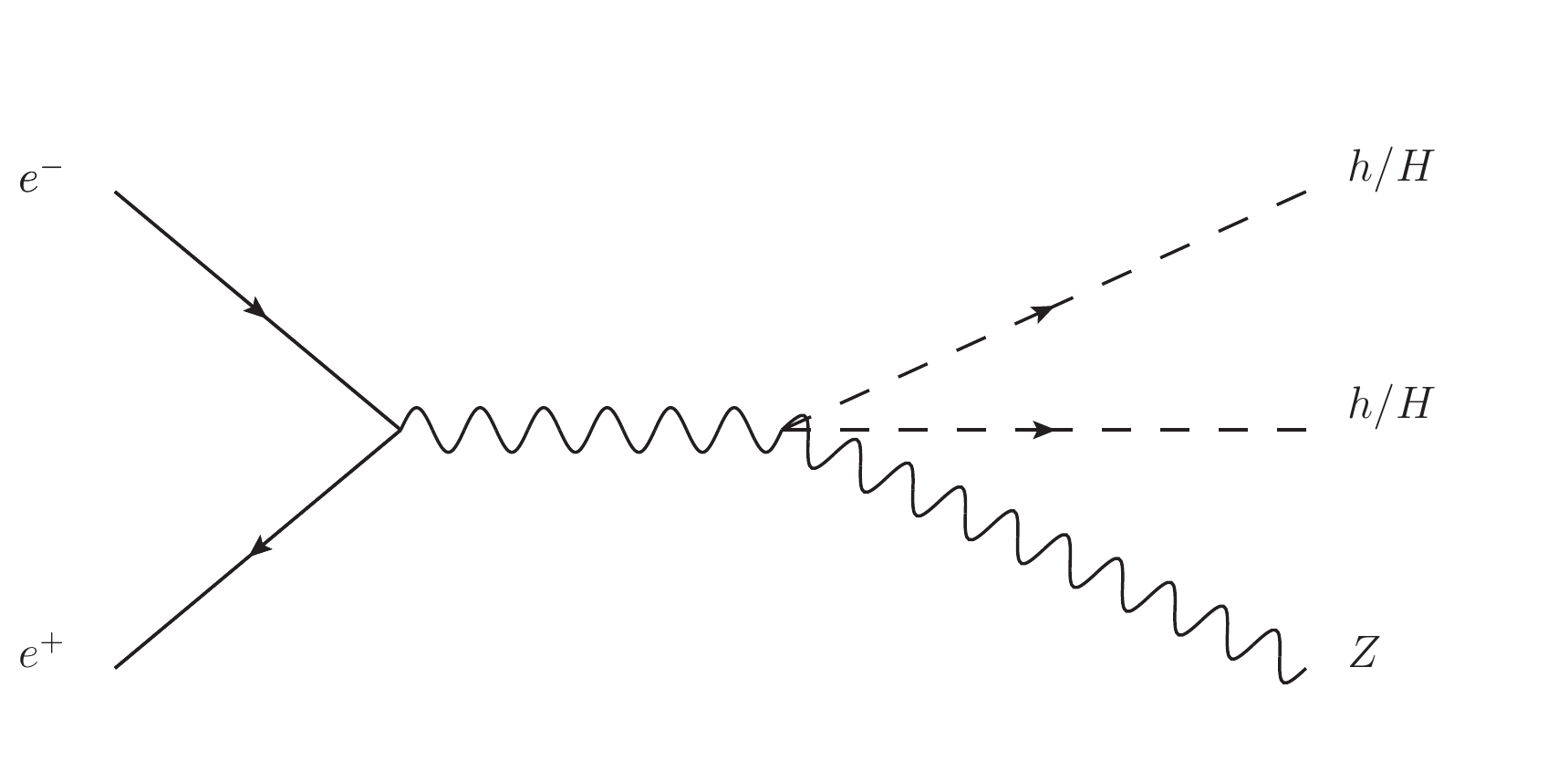}
\caption{The Feynman diagrams for the Higgs boson pair productions associated with a $Z$-boson at the ILC. }
\label{fig:hhZ_diagrams}
\end{figure}

The ILC can directly produce Higgs boson pairs associated with a $Z$-boson, when it runs at the center-of-mass energy of $\sqrt{s}=500\,\GeV$.
The Feynman diagrams for the corresponding processes are depicted in Fig.~\ref{fig:hhZ_diagrams}.
The cross sections at the ILC can be expressed as~\cite{Hikasa:1985qi}
\beqn\label{eq:hhZ_ILC}
\sigma&=& \frac{1}{4} \Big[  (1 + P_{e^-} )(1+P_{e^+} ) \sigma_{\rm RR} + (1 - P_{e^-} )(1 - P_{e^+} ) \sigma_{\rm LL} \non
&+&  (1 + P_{e^-} )(1 - P_{e^+} ) \sigma_{\rm RL} + (1 - P_{e^-} )(1 + P_{e^+} ) \sigma_{\rm LR}   \Big]\,,
\eeqn
where $\sigma_{\rm LR}$ denotes the cross section at beam polarization configurations of $(P_{e^+}\,, P_{e^-})=(+1\,, -1)$, and etc. 
The ILC will run at $\sqrt{s}=500\,\GeV$ with an integrated luminosity of $4\,\ab^{-1}$, which will be equally shared by two beam polarization configurations of $(P_{e^+}\,, P_{e^-})=(\pm 0.3\,, \mp 0.8)$~\cite{Fujii:2015jha, Barklow:2015tja}. 
The corresponding cross sections for the $e^+ e^- \to hhZ$ read
\beqs
\beqn
\sigma_{(+0.3\,, -0.8)}&=& 0.585\, \sigma_{\rm LR} + 0.035\, \sigma_{\rm RL}\,,\\
\sigma_{(-0.3\,, +0.8)}&=& 0.035\, \sigma_{\rm LR} + 0.585\, \sigma_{\rm RL}\,.
\eeqn
\eeqs

\begin{table}[!hbt]
\centering
\begin{tabular}{cccc}
\hline
\hline
$P(e^+,e^-)$ & Channel & Excess significance & Precision on $\sigma_{ZHH}$ \\
\hline
(0.3,-0.8) & $HH\to b\bar bb \bar b$ & $3.5\,\sigma$ & 30.3\% \\
(-0.3,0.8) & $HH\to b\bar bb \bar b$ &  $4.8\,\sigma$ & 29.4\% \\
(0.0,-0.8) & $HH\to b\bar bb \bar b$ & $3.5\,\sigma$ & 34.7\% \\
(0.0,0.8)  & $HH\to b\bar bb \bar b$ & $4.2\,\sigma$ & 33.7\% \\
(0.6,-0.8) & $HH\to b\bar bb \bar b$ & $4.2\,\sigma$ & 28.7\% \\
(-0.6,0.8) & $HH\to b\bar bb \bar b$ & $5.5\,\sigma$ & 27.8\% \\
\hline
(0.3,-0.8) & $HH\to b \bar b W^+  W^-$ & $1.91\,\sigma$ & ... \\
\hline
\hline
\end{tabular}
\caption{
The prospects of measuring the Higgs pair productions at the ILC. 
For each beam polarization configuration, an integrated luminosity of $\mL = 2\text{ ab}^{-1}$ is assumed.}
\label{tab:ILCZHH}
\end{table}

\begin{figure}[!thb]
\centering
\includegraphics[width=0.70\textwidth,trim=10 0 0 10]{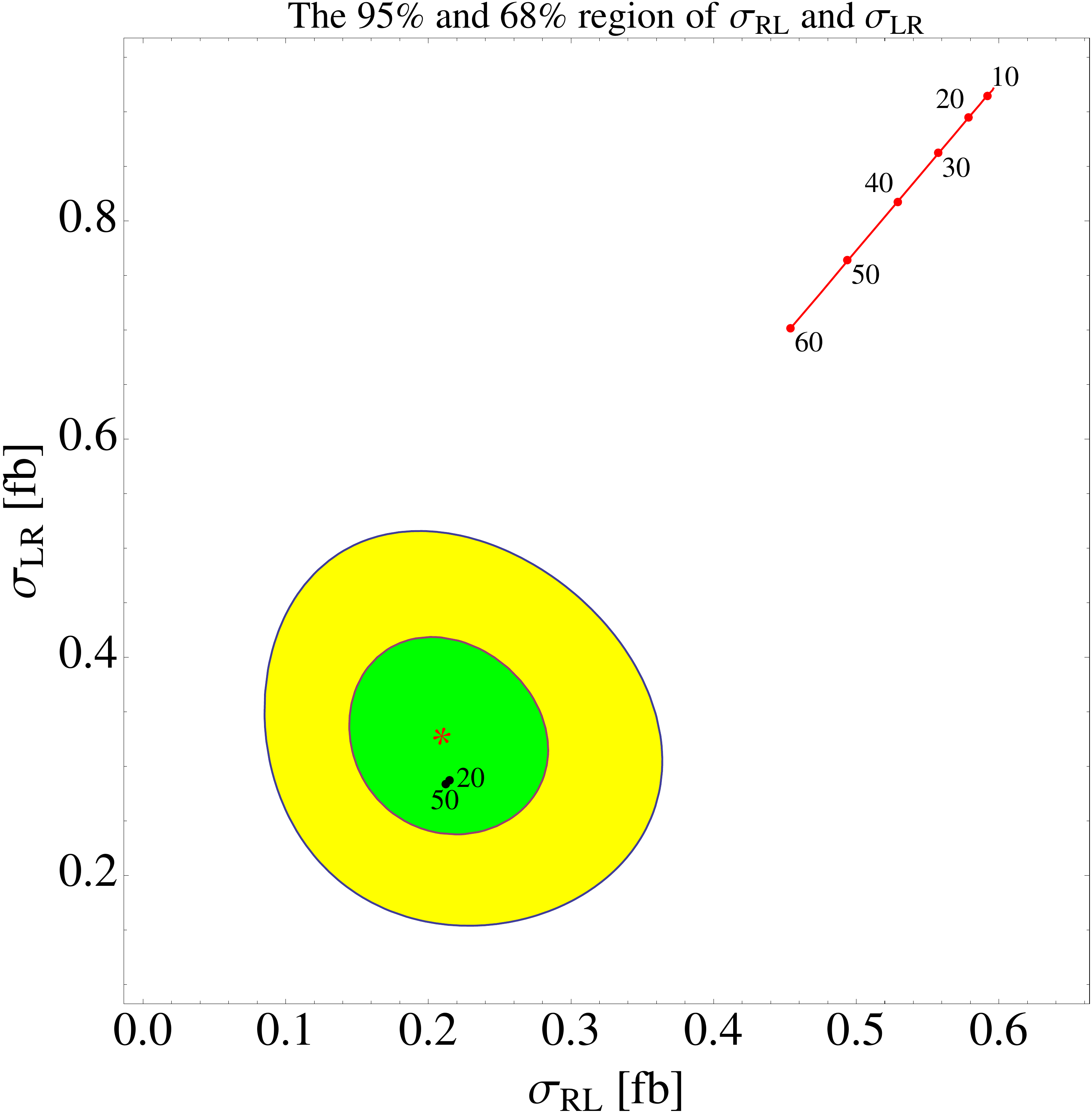}
\caption{
95\% and 68\% contours of the measurement of $\sigma_{ZHH,LR}$ and $\sigma_{ZHH,RL}$ from (1) $hh\to b \bar b b \bar b$ with $P(e^-,e^+)=(\pm0.8,\mp0.3)$~\cite{Duerig:2016dvi} and (2) $hh\to b \bar b W^+ W^-$ with $P(e^-,e^+)=(-0.8,0.3)$~\cite{LC-REP-2013-025} for 500 GeV ILC with 2 ab$^{-1}$ luminosity for each beam polarization configuration. 
The best-fit points of $(\alpha\,,\beta)=(0.04\,,1.40)$ for Type-I case and $(\alpha\,,\beta)=(-0.25\,, 1.31)$ for Type-II case are taken as in Table.~\ref{tab:BM_hH}.
The tiny black line represents the prediction of the cross section (normalized according to the branching fraction) of Type-I with variation of $m_{12}$, the red line is for Type-II. 
The label beside each point indicates the value of corresponding $m_{12}$.}
\label{fig:PrecisionSigmaZHH}
\end{figure}

\begin{figure}[!thb]
\centering
\includegraphics[width=0.49\textwidth,trim=30 0 18 0]{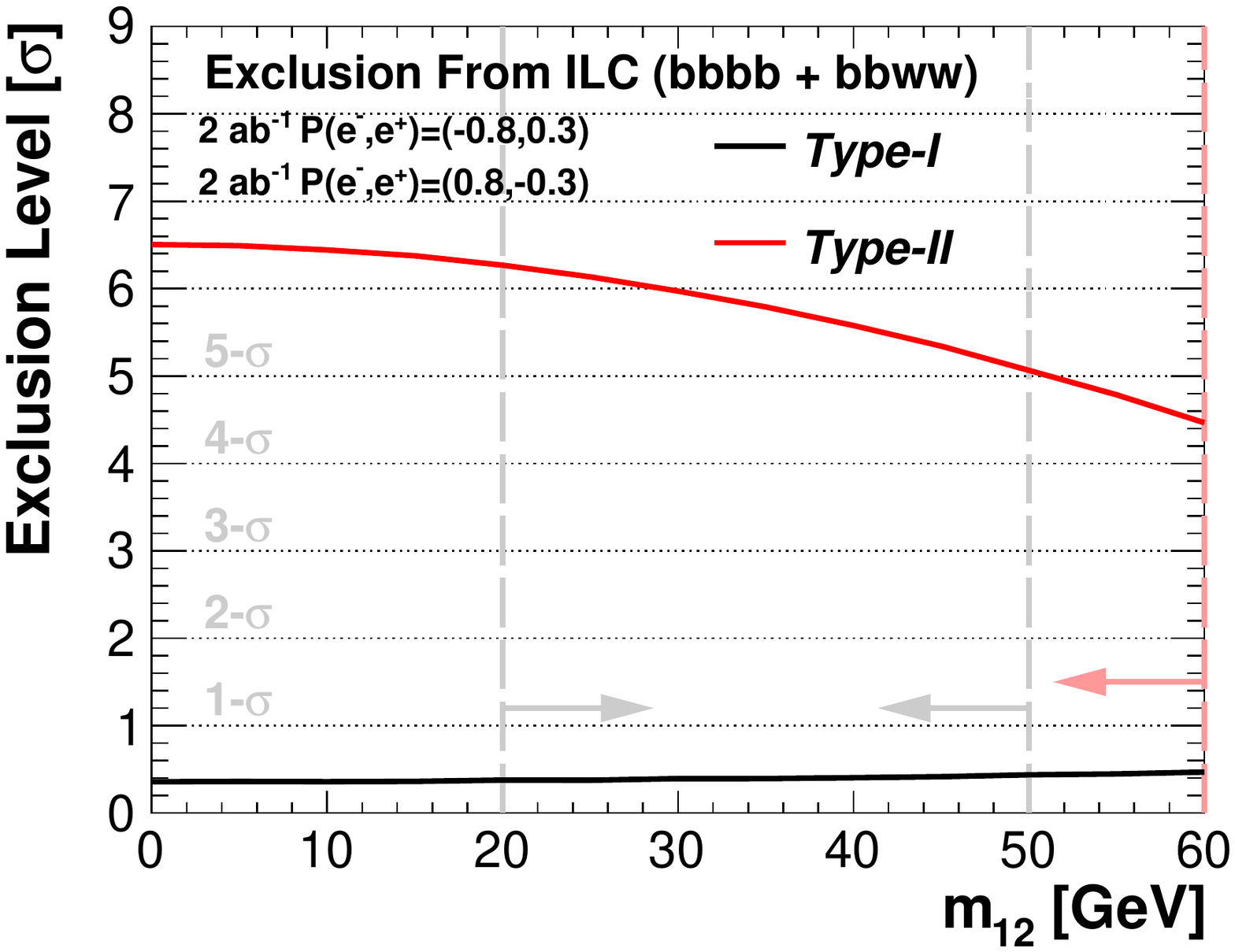}
\includegraphics[width=0.49\textwidth,trim=30 0 18 0]{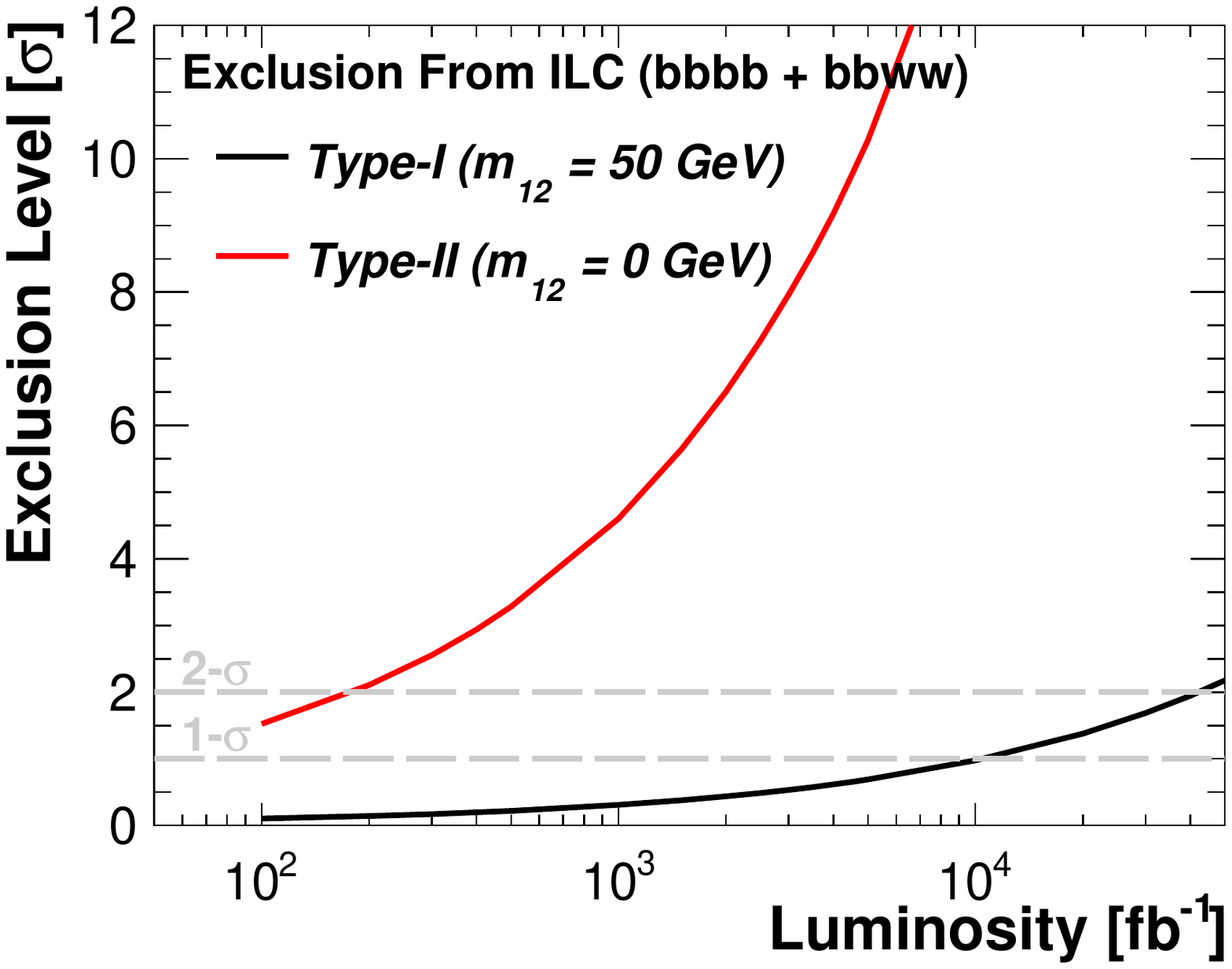}
\caption{
Left: the exclusion level of the hypothesis test of the 2HDM model against SM using $hh\to bbbb$ and $hh\to bbWW$ channels with $(P_{e^+}\,, P_{e^-})=(\pm 0.3\,, \mp 0.8)$ and 2 ab$^{-1}$ luminosity for each polarization states. The black (red) line represents the exclusion level for the $h/H$ degenerate in the Type-I (Type-II) 2HDM with dependence on $m_{12}$.
The gray and red vertical dashed lines with arrow directions in the left panel indicate the allowed upper bounds from the 2HDM potential stability constraints for Type-I and Type-II, respectively. 
Right: the exclusion level the same as in the left panel, but with dependence on the luminosity cumulated for each beam polarization configuration, $m_{12} = 50\,(0)$ GeV is used for Type-I (Type-II) for presentation.}
\label{fig:ExclusionM12}
\end{figure}

Prospects of measuring the cross sections of the SM Higgs boson pair production via the $hh\to b \bar bb \bar b$ and $hh\to b \bar b W^+ W^-$ final states have been investigated in Refs.~\cite{Duerig:2016dvi,LC-REP-2013-025}. 
The corresponding significance are listed in Table.~\ref{tab:ILCZHH}. 
Combining these measurements for $(P_{e^+}\,, P_{e^-})=(\pm 0.3\,, \mp 0.8)$, we obtain the ILC precision on the measurement of $(\sigma_{\rm RL}\,,\sigma_{\rm LR})$ in Fig.~\ref{fig:PrecisionSigmaZHH}, with two contours for 1- and 2-$\sigma$ regions, respectively. 
In Fig.~\ref{fig:PrecisionSigmaZHH}, we also show the best-fit points for Type-I (Black) and Type-II (Red) 2HDM in the same plane.
Each point represents different value of $m_{12}$ given by the corresponding label. 
Note that the 2HDM cross sections are calculated by using the best-fit points in the $(\alpha\,,\beta)$ plane from the LHC global fitting in previous section.
From Fig.~\ref{fig:PrecisionSigmaZHH}, we find that for the $h/H$ mass-degenerate Type-I 2HDM case, the cross sections for all allowed $m_{12}$ are still within the precision of the ILC measurement.
However, the $h/H$ mass-degenerate Type-II 2HDM points can be excluded from the Higgs pair production measurements at the ILC.

More details can be found in Fig.~\ref{fig:ExclusionM12}, where we combine $hh\to b \bar b b \bar b$ and $hh\to b \bar b W^+ W^-$ channels and use the log-likelihood ratio method to perform the hypothesis test against the SM predictions. 
The exclusion levels for different $m_{12}$ values are presented by the black and red line in the left panel for Type-I and Type-II cases, respectively. 
Dashed vertical lines with arrows indicate the allowed region from 2HDM potential unitarity/stability constraints. 
We find that for Type-II case, the theoretical allowed region of $m_{12}$ has already been excluded by the ILC measurement at $4\,\sigma$ or more, with $2\,\ab^{-1}$ luminosity for each beam polarization configuration. 
In contrast, the allowed region for the Type-I case is still safely sitting within $1\,\sigma$ region of the ILC measurement. 
In the right panel, we present the exclusion level as a function of the luminosity cumulated at the ILC for beam polarization configurations of $P(e^-,e^+)=(\pm0.8,\mp0.3)$. 
The black and red line are for Type-I and Type-II case, respectively. 
For illustration, we choose the most sensitive value of $m_{12}$ in the stability allowed region for each type: $m_{12} = 50\,(0)\,\GeV$ for Type-I (Type-II). 
We can see that, the ILC has sensitivity to Type-II in this case when an integrated luminosity of $400\,\fb^{-1}$ (equally shared by two beam polarization configurations) is cumulated. 
However, at least an integrated luminosity of $80\,\ab^{-1}$ is required for the ILC to be sensitive to the Type-I in this situation,  which is unrealistic.


\section{Conclusion}
\label{section:conclusion}

In this work, we study the future prospects of distinguishing the mass-degenerate Higgs boson scenario from the single resonance case at the LHC and future $e^+ e^-$ colliders.
Our study is performed in the general CPC 2HDM framework, with two $CP$-even Higgs bosons of $h/H$ to be mass-degenerate.
The direct measurements of the Higgs boson signal rates at the $125\,\GeV$ only probe its/their gauge couplings and Yukawa couplings.
Alternatively, we find this scenario can be further constrained by a series of theoretical bounds and direct experimental searches in such a framework.
Moreover, we suggest that the study of the Higgs boson pair productions will be useful for this scenario.
Specifically, there are four types of Higgs cubic self couplings involved in the Higgs pair productions, which are $(\lambda_{hhh}\,, \lambda_{hhH}\,, \lambda_{hHH}\,, \lambda_{HHH} )$.
The physical processes to be considered are the ggF to Higgs boson pair productions at the LHC, and the Higgs boson pair productions associated with a $Z$-boson at the ILC.

By performing the global fit of the LHC measurements of the $125\,\GeV$ Higgs boson signal rates, we find the best-fit points with mass-degenerate $h/H$ in Type-I and Type-II 2HDM in $(\alpha\,,\beta)$ plane.
The best-fit point in the Type-I case deviate from the alignment limit as large as $c_{\beta-\alpha}\simeq 0.21$, while it approaches to the alignment limit as $c_{\beta-\alpha}\sim \mO(10^{-2})$ in the Type-II case.
Correspondingly, one Higgs boson $H$ becomes almost gaugephobic in the Type-II case.
The $h/H$ mass-degenerate scenario also passes the current LHC run-II searches for the $CP$-odd Higgs boson $A$ via the $b \bar b +\ell^+ \ell^-$ final states. 
Meanwhile, it also means that the further LHC searches for the $CP$-odd Higgs boson $A$ below the $t \bar t$ mass threshold will play a role to justify or falsify the $h/H$ mass-degenerate scenario.

The relevant cubic Higgs self couplings, such as $\lambda_{hHH}$ and $\lambda_{HHH}$, are not vanishing even when the 2HDM parameters approach to the alignment limit.
This suggests that the Higgs pair production processes can crucial to justify or falsify the $h/H$ mass-degenerate scenario.
The signal predictions of the ggF to mass-degenerate Higgs boson pairs are made at the LHC $14\,\TeV$ runs, with the focus on two leading search channels of $4b$ and $2b\,2\gamma$.
Moderate signal enhancements with respect to the SM predictions are expected, while the enhancements are at most $\sim 10\,\%$. 
Therefore, the Higgs boson pair productions at the LHC are less likely to probe the $h/H$ mass-degenerate scenario.
At the ILC $500\,\GeV$ run, we find that the $h/H$ mass-degenerate samples in the Type-I case are within the precision of the ILC measurement, while the $h/H$ mass-degenerate samples in the Type-II case can be probed or excluded with an integrated luminosity of $400\,\fb^{-1}$.
It means that the ILC $500\,\GeV$ run offers an opportunity to fully probe the $h/H$ mass-degenerate scenario in the Type-II case.

Though our predictions are model-dependent, the Higgs pair productions with two mass-degenerate Higgs bosons can be generalized to any other NP models with multiple Higgs bosons.
There are multiple Higgs self couplings involved in the Higgs pair productions in general.
Depending on the model setup, these self-couplings may be bounded by the constraints mentioned in the current study.
Our discussion through the context focus on the ggF process at the LHC and the Higgs pair strahlung at the ILC.
This discussion can be extended to other Higgs pair production channels, including the vector boson fusion (VBF) and $t\bar t$ associated processes at the future high-energy $e^+ e^-$ and $pp$ colliders.


\section*{ACKNOWLEDGMENTS}

We would like to thank Yanwen Liu, Zuowei Liu, Manqi Ruan, and Junping Tian for very useful discussions and communication. 
The work of N.C. is partially supported by the National Natural Science Foundation of China (under Grant No. 11575176) and Center for Future High Energy Physics (CFHEP). 
The work of L.G.B. is partially supported by the National Natural Science Foundation of China (under Grant No. 11605016), Basic Science Research Program through the National Research Foundation of Korea (NRF) funded by the Ministry of Education, Science and Technology (NRF-2016R1A2B4008759), and Korea Research Fellowship Program through the National Research Foundation of Korea (NRF) funded by the Ministry of Science and ICT (2017H1D3A1A01014046). 
The work of Y.C.W is partially supported by the Natural Sciences and Engineering Research Council of Canada.
The work of Y.Z. is partially supported by the National Natural Science Foundation of China (under Grant Nos. of 11775109 and U1738134), by he Nanjing University (under Grant No.14902303), the China Postdoctoral Science Foundation (under Grant No. 2017M611771), and the Fundamental Research Funds for the Central Universities (under Grant No. 020414380097).
N.C. would like to thank Nanjing University for their hospitalities when part of this work was prepared.

\newpage

\bibliographystyle{unsrt}
\bibliography{references}

\end{document}